\documentclass[11pt]{article}

\usepackage[a4paper,top=2.2cm,bottom=2.2cm,left=2.2cm,right=2.2cm,marginparwidth=1.75cm]{geometry}
\usepackage{color}
\usepackage{hyperref}
\hypersetup{
     colorlinks = true,
     linkcolor = blue,
     anchorcolor = blue,
     citecolor = blue,
     filecolor = blue,
     urlcolor = blue
     }
\usepackage[authoryear]{natbib}
\setcitestyle{authoryear,yysep={,}, notesep={, },round, comma, aysep={;}}
\usepackage{amsmath}
\usepackage{amssymb}
\usepackage{amsfonts}
\usepackage{epsfig}
\usepackage{graphicx}
\usepackage{graphics}
\usepackage{float}
\usepackage{eurosym}
\usepackage{enumitem}
\usepackage{multirow}
\usepackage{caption}
\usepackage{subcaption}
\usepackage{xspace}
\makeatletter\DeclareRobustCommand\onedot{\futurelet\@let@token\@onedot}
\def\@onedot{\ifx\@let@token.\else.\null\fi\xspace}
\def\etal{\emph{et al}\onedot}

\DeclareMathOperator*{\argmin}{argmin}

\newtheorem{definition}{Definition}
\newtheorem{assumption}{Assumption}
\newtheorem{property}{Property}
\newtheorem{corollary}{Corollary}
\newtheorem{proposition}{Proposition}
\newtheorem{theorem}{Theorem}

\newtheorem{example}{Example}[section]

\begin{document}

\title{\huge {\bf Regression Markets and Application to Energy Forecasting} \vspace{1cm}}
 
\author{\Large Pierre Pinson$^{\dag *}$, Liyang Han$^\ddag$, Jalal Kazempour$^\ddag$ \\ \textcolor{white}{.} \\ \small $^\dag$ Technical University of Denmark \\ \small Department of Technology, Management and Economics \\ \small $^*$ email: \href{mailto:ppin@dtu.dk}{ppin@dtu.dk} \\
\textcolor{white}{.} \\ \small $^\ddag$ Technical University of Denmark \\ \small Department of Electrical Engineering \\ \small Centre for Electric Power and Energy \vspace{1cm}}

\date{\today}

\maketitle

\begin{abstract}
Energy forecasting has attracted enormous attention over the last few decades, with novel proposals related to the use of heterogeneous data sources, probabilistic forecasting, online learning, etc. A key aspect that emerged is that learning and forecasting may highly benefit from distributed data, though not only in the geographical sense. That is, various agents collect and own data that may be useful to others. In contrast to recent proposals that look into distributed and privacy-preserving learning (incentive-free), we explore here a framework called regression markets. There, agents aiming to improve their forecasts post a regression task, for which other agents may contribute by sharing their data for their features and get monetarily rewarded for it. The market design is for regression models that are linear in their parameters, and possibly separable, with estimation performed based on either batch or online learning. Both in-sample and out-of-sample aspects are considered, with markets for fitting models in-sample, and then for improving genuine forecasts out-of-sample. Such regression markets rely on recent concepts within interpretability of machine learning approaches and cooperative game theory, with Shapley additive explanations. Besides introducing the market design and proving its desirable properties, application results are shown based on simulation studies (to highlight the salient features of the proposal) and with real-world case studies.

\vspace{1cm}
\end{abstract}

\newpage

\section{Introduction}
\label{sec:intro}

Renewable energy forecasting has evolved tremendously over the last 10-20 years, with a strong evolution towards probabilistic forecasting, cutting-edge statistical and machine learning approaches, the use of large amounts of heterogeneous and distributed data, etc. For a recent and compact review of the state of the art withing energy forecasting, the reader is referred to \citet{Hong2020}. \textcolor{black}{Especially when it comes to the use of heterogeneous and distributed data sources, numerous works support the idea that forecasting quality may be substantially improved, see \citet{Andrade2017}, \citet{Cavalcante2017} and \citet{Messner2019} among others. These works have shown that improvements may be obtained by using offsite information (e.g., power and meteorological measurements) as well as weather forecasts over neighboring grid points, for areas covering tens to a few hundreds of kilometers. Improvements are observed for forecasts in the form of conditional expectations, but also for probabilistic forecasts, e.g., quantiles, intervals and predictive densities.} When using the term {\it distributed}, we here mean both in the geographical and ownership sense, i.e., the data potentially valuable to a given agent of the energy system is actually collected and owned by other agents. Therefore, some have pushed forward proposals towards distributed and privacy-preserving learning \citep{Zhang2018, Sommer2021}, as a way to get the benefits from such distributed data, without revealing the private information of the agents involved. Beyond energy applications, this approach is generally known as federated learning \citep{Li2020}, with substantial developments over the last few years. The alternative that we propose to explore here is that of \textit{data monetization} within a collaborative and market-based analytics framework. \textcolor{black}{In the frame of the paper, it is assumed that if remunerated, agents are willing to share their actual data with an analytics platform. Privacy-related aspects are hence not readily considered, since data is shared with the platform but not with other agents. If privacy was to be additionally accommodated in that collaborative and market-based analytics framework, alternative approaches relying on distributed computing, differential privacy, etc. could be employed. As a representative example, \citet{Goncalves2021} analyzed some of these alternative approaches in a collaborative forecasting context.}

Concepts of information sharing have been prevalent in some parts of the economics and game-theory focused literature, going as far back as the 1980s \citep{Gal-Or1985}. Data monetization and data markets have been increasingly discussed over the last 5-10 years, with a number of proposals towards algorithmic solutions \citep{Agarwal2019}, as well as fundamental aspects of pricing and privacy-preservation \citep{Acemoglou2019}, more generally also with consideration of bilateral exchange of data vs.\ monetization of data \citep{Rasouli2021}. For a recent review of the state of the art related to data markets, see \citet{Bergemann2019} and \citet{Liang2018}. \textcolor{black}{Approaches that would be suitable for renewable energy forecasting and energy applications more broadly are scarse though, with the notable recent example of \citet{Goncalves2020}, who adapt and apply an approach in line with the proposal of \citet{Agarwal2019}, restricted to batch learning and in-sample assessment of the value of data and features provided. Renewable energy forecasting appears to be an ideal playground to develop, apply and assess data markets, in view of the known value of distributed data, the liberalization of energy markets, and potential resulting impact. In addition, such data markets can then be developed along the lines of cutting-edge forecasting frameworks, where forecasts are thought of within a probabilistic framework, the environment is seen as nonstationary, etc. As of today, there does not exist such data markets that would jointly look at in-sample and out-of-sample value of data for forecasting, as well as both batch and online learning in underlying regression models.} Consequently here, our aim is to describe and to analyze a theoretically sound and practical proposal for data monetization within a collaborative and market-based analytics framework, which is readily suitable for energy-related forecasting applications and these aforementioned characteristics. We restrict ourselves to a market with a single buyer and multiple sellers. This corresponds to the case where an agent that would like to improve the solving of a regression task posts this task on an analytics platform, where other agents can come and propose their features and own data. Many tasks could be posted in parallel, but buyers or tasks would not compete for the features and data to be supplied. However, several tasks could be posted and handled in parallel (as in our case-study application) based on the idea that buying the data does not bring exclusivity. Exclusivity is here defined as the fact that if data is sold to an agent, it cannot be sold to another agent in parallel. In contrast, if aiming for exclusivity, other setups exist for feature allocation among multiple buyers and sellers with the aim of maximizing social welfare, as for the example case of \citet{Cao2017}.

Within energy forecasting applications, one most often finds a regression model and a learning process used to fit model parameters. Therefore, we place our focus on so-called {\it regression markets}. These markets readily build on the seminal work of \citet{Dekel2010}, who were the first to look at mechanism design aspects for a regression setting where agents may be strategic in the way they share private information. Here, regression markets are considered in both batch and online versions, since modern learning and forecasting techniques mostly rely on these two approaches. We restrict ourselves to a certain class of regression problems (linear in parameters), which allows us to obtain certain market properties. It was already shown and discussed by \citet{Dekel2010} that certain properties, especially truthfulness (also referred to as incentive compatibility) is difficult to obtain in a more general regression setting. Extensions to privacy-constrained truthful regression, limited to a linear setting, were also recently discussed \citep{Cummings2015}. The quality of the model fitting is assessed by a negatively-oriented convex loss function $l$ (lower is better), which may be quadratic in the case of Least-Squares (LS) fitting, a smooth quantile loss in the case of quantile regression, a Maximum Likelihood (ML) score for more general probabilistic models, etc. That convex loss function is at the core of our proposal, since the main idea is that an agent may be able to decrease the loss $l$ by using data from other agents. These agents should be monetarily compensated in a fair and efficient way, i.e., in line with their individual and marginal contribution to improvements in $l$. For that purpose, we use some recent concepts related to interpretability in machine learning, following the original proposal of \citet{Lundberg2017} and the wealth of subsequent proposals, which directly connect to a cooperative game-theoretical framework as in \cite{Agarwal2019}. Finally, a particular aspect of our contribution is that we consider both in-sample (i.e., model fitting based on past data) and out-of-sample (i.e., use of those models for forecasting based on new data) since, in actual energy forecasting application, both need to be considered in order to improve model fitting, but also genuine forecast quality.

The document is structured as following: firstly, Section~\ref{sec:setup} describes the agents and preliminaries regarding regression tasks. Subsequently, Section~\ref{sec:regmarket} introduces our original proposal for regression market mechanisms, where agents are monetarily rewarded for their contribution to improving the solving of a given regression task, in the sense of lowering the convex loss function $l$. The overall concept is presented for both batch and online setups, also with a description of feature valuation and allocation policies. The extension to the out-of-sample regression and forecasting case is also covered. The properties of our regression market mechanisms are finally presented and proven. The approach is illustrated based on a set of simulation studies, which are gathered in Section~\ref{sec:simulation}, for a broad range of models and cases. Section~\ref{sec:appl} then describes and discusses an application to real-world forecasting case-studies, with both mean and quantile forecasting problems, as well as batch and online learning. Finally, Section~\ref{sec:concl} gathers a set of conclusions and perspectives for future work.

\section{Setup, regression and estimation}
\label{sec:setup}

\subsection{Central and support agents}

Consider a set of agents $\mathcal{A} = \{a_1,a_2,\hdots,a_m\}$. Out of this set of agents, one of the agents $a_i \in \mathcal{A}$ is referred to as \emph{central agent}, in the sense that this agent has an analytics task at hand, in the form of a regression problem for an eventual forecasting application. We refer to the other agents $a_j, \, j\neq i$, as {\it support agents}, since they may be supporting the central agent with the analytics task at hand. The central agent has a target variable $\{Y_t\}$, seen as a stochastic process, i.e., a succession of random variables $Y_t$ indexed over time, with $t$ the time index. Eventually, a time-series $\{y_t\}$ is observed, which consists of realizations from $\{Y_t\}$, one per time index value. For simplicity, we consider that realizations of $Y_t$ can take any value in $\mathbb{R}$, even though in practice, it is also fine if restricted to a subset of it (positive values only, or within the unit interval $[0,1]$, for instance).

The central agent aims at obtaining a model that can describe some given characteristics $z_t$ of $Y_t$, e.g., its mean $\mu_t$ or a specific quantile $q^{(\tau)}_t$ with nominal level $\tau$. This description relies on a set $\Omega = \{x_k, \, k=1,\hdots,K\}$ of input features (also referred to as explanatory variables). These features and their observations are distributed among all agents. We denote by $x_{k,t}$ the observation of feature $x_k$ at time $t$. As for the target variable, we consider for simplicity that $x_{k,t} \in \mathbb{R},\, \forall t,k$, though in practice these may also be restricted to a subset of $\mathbb{R}$. 

All features and target variable are observed at successive time instants, $t=1,\hdots,T$, such that we eventually have time-series of those. Let us write $\mathbf{x}_k = [x_{k,1} \, \hdots \, x_{k,T}]^\top$ the vector of values for the feature $x_k$, $\mathbf{x}_t = [x_{1,t} \, \hdots \, x_{K,t}]^\top$ the vector of values for all features at time $t$, while $\mathbf{y} = [y_1 \, \hdots \, y_T]^\top$ gathers all target variable observations, over the $T$ time steps. In the case only a subset of features $\omega \subset \Omega$ is used, the vector of feature values at time $t$ is denoted by $\mathbf{x}_{\omega,t}$. In practice such features may be observations (meteorological, power measurements, etc.) or forecasts (e.g., for weather variables). We write $\mathbf{X}_\omega \in \mathbb{R}^{T\times |\omega|}$ the design matrix, the $t^{\text{th}}$ row of which is $\mathbf{x}_{\omega,t}^\top$.

The features are distributed among all agents in $\mathcal{A}$ as following: the central agent $a_i$ owns a set $\omega_i$ (of cardinal $|\omega_i|$) of features, $\omega_i \subset \Omega$, as well as the target variable $y$; the support agents, gathered in the set $\mathcal{A}_{- i} = \{a_j, \, j \neq i\}$, own the other input features, which could be of relevance to the central agent for that regression task. Each agent $a_j$ has a set $\omega_j$ with $|\omega_j|$ features, $\omega_j \subset \Omega$, such that $|\omega_i|+ \sum_j |\omega_j| = K$. We write $\Omega_{-i}$ the set that contains the features of support agents only, $\Omega_{-i} = \Omega \setminus \omega_i$ .

\subsection{Regression framework}

\subsubsection{Regression models that are linear in their parameters}

Generally speaking, based on temporally index data, collected at regular time intervals, a regression problem aims at describing the mapping $f$ between a set $\omega \subset \Omega$ of explanatory variables, and the target variable $z$, i.e.,
\begin{equation}\label{eq:regression}
 f \, :  \mathbf{x}_{\omega,t} \in \mathbb{R}^{\left|\omega \right|} \rightarrow z_t \in \mathbb{R} \, .
\end{equation}
In principle $f$ may be linear or nonlinear, and a wealth of approaches can be considered for its modeling. We restrict ourselves to the case of parametric regression in the sense that
\begin{equation}\label{eq:regressionadd}
 z_t = f(\mathbf{x}_{\omega,t};\boldsymbol{\beta}_\omega) \, , \quad \forall t \, .
\end{equation}
Consequently, given a structural choice for $f$, the regression may be fully and uniquely described by the set of parameters $\boldsymbol{\beta}_\omega = [\beta_0 \, \, \beta_1 \, \, \hdots \, \, \beta_n]^\top$, $n \geq \left|\omega \right|+1$. In the linear regression case, $n=\left|\omega \right|+1$, while $n > \left|\omega \right|+1$ for nonlinear regression. We additionally restrict ourselves to the case of regression models that can be expressed as linear in their parameters $\boldsymbol{\beta}_\omega$, since if using convex loss functions the resulting estimation problem is convex too. That class of regression problems is not limited to linear regression only though, since also covering nonlinear regression problems such as polynomial regression, local polynomial regression, additive models with splines, etc. This therefore means the model in \eqref{eq:regressionadd} can be expressed as
\begin{equation}\label{eq:regressionaugm}
 z_t = \boldsymbol{\beta}_\omega^\top \tilde{\mathbf{x}}_{\omega,t} \, , \quad \forall t \, ,
\end{equation}
where $\tilde{\mathbf{x}}_{\omega,t} \in \mathbb{R}^n$ is the observation at time $t$ of the augmented feature vector $\tilde{\mathbf{x}}_{\omega}$. For instance if having $K=2$ features $x_1$ and $x_2$ and considering polynomial regression of order 2, the augmented feature vector at that time can written as $\tilde{\mathbf{x}}_{\omega,t} = [1 \, \, x_{1,t} \, \, x_{2,t} \, \, x_{1,t}^2 \, \, x_{1,t} x_{2,t} \, \, x_{2,t}^2]$. The vector of parameters $\boldsymbol{\beta}_\omega$ hence has dimension $n=6$.

In the following, to place ourselves in the most generic framework, we focus on the regression problems as in \eqref{eq:regressionaugm}, as they also encompass basic linear regression when $\tilde{\mathbf{x}}_\omega = \mathbf{x}_\omega$. We write $\tilde{\mathbf{X}}_\omega \in \mathbb{R}^{T\times n}$ the design matrix, the $t^{\text{th}}$ row of which is $\tilde{\mathbf{x}}_{\omega,t}^\top$.

\subsubsection{Separable and non-separable regression problems}

Consider the general case for which a linear regression model $f$ uses features $x_k$ within a set $\omega \subset \Omega$ as input (so, possibly from both central and support agents), to describe a characteristic $z_t$ of $Y_t$. Linear regression relies on the following model for $Y_t$,
\begin{equation} \label{eq:lingeneral}
    Y_t = \beta_0 + \sum_{k|x_k \in \omega} \beta_k x_{k,t} + \varepsilon_t \, , \quad \forall t \, ,
\end{equation}
where $\varepsilon_t$ is a centred noise with finite variance. This readily translates to
\begin{equation} \label{eq:lingeneral2}
    z_t = \beta_0 + \sum_{k|x_k \in \omega} \beta_k x_{k,t} \, , \quad \forall t \, .
\end{equation}
For instance, if $z_t$ is the expectation of $Y_t$, this means that this expectation is modeled as a linear function of the input feature values at time $t$.

In the special case where only the features of the central agent $a_i$ are used, one has
\begin{equation} \label{eq:lincentral}
    z_t = \beta_0 + \sum_{k|x_k \in \omega_i} \beta_k x_{k,t} \, , \quad \forall t \, ,
\end{equation}
i.e., only considering the features owned by the central agent, $x_k \in \omega_i$. The $\beta_k$'s are hence the coefficients in the linear model corresponding to the features owned by the central agent. In contrast, if the features of all support agents were also considered, the corresponding linear model would be
\begin{equation} \label{eq:linall}
    z_t  = \beta_0 + \sum_{k=1}^{K} \beta_k x_{k,t} = \underbrace{\beta_0 + \sum_{k|x_k \in \omega_i} \beta_k x_{k,t}}_{\text{central agent}} \, + \underbrace{\sum_{j \in \mathcal{A}_{-i}} \sum_{k|x_k \in \omega_{j}} \beta_k x_{k,t}}_{\text{support agents}}  \, , \quad \forall t \, ,
\end{equation}
where the $\beta_k$'s (related to $x_k \in \Omega_{-i}$) are the coefficients in the linear model corresponding to the features owned by the all support agents. In principle, $\beta_0$ could be taken aside since not relating to a feature owned by neither central nor support agents. For simplicity in the following, we consider that the central agent also has a unit feature, that hence corresponds to that intercept. As can be seen from \eqref{eq:linall}, such linear regression models are separable, in the sense that we can separate blocks of terms that relate to the individual features of each agent. Similarly, additive models with splines are separable, since these may be written as
\begin{equation} \label{eq:addlinall}
    z_t  = \beta_0 + \sum_{k=1}^{K} g_k(x_{k,t}) = \underbrace{\beta_0 + \sum_{k|x_k \in \omega_i} g_k(x_{k,t})}_{\text{central agent}} \, + \underbrace{\sum_{j \in \mathcal{A}_{-i}} \sum_{k|x_k \in \omega_{j}} g_k(x_{k,t})}_{\text{support agents}}  \, , \quad \forall t \, ,
\end{equation}
where
\begin{equation} \label{eq:splinebasis}
    g_k(x_{k,t}) = \sum_{i=1}^{n_k} \beta_{k,i} B_i (x_{k,t}) \, , \quad \forall k,t \, .
\end{equation}
In the above, the $B_i$'s denotes the basis functions, while $g_k$ is the spline basis expansion relying on $n_k$ basis functions. In addition, $n_k$ is the number of degrees of freedom, being itself a function of spline type and the number of knots. By combining~\eqref{eq:addlinall} and~\eqref{eq:splinebasis}, one sees that additive models with a spline basis take the form of the generic parametric regression model~\eqref{eq:regressionaugm}, and that these are separable.

In contrast, if using polynomial regression (as well as local polynomial regression) with degree greater than 1, the regression models are not separable, since interaction terms in the form of direct multiplication of features owned by different agents will be present. Consequently, one cannot have this separation in blocks as for  linear regression and additive models with splines. To illustrate those situations, two examples are gathered below.

\begin{example}[ARX model for the mean] The central agent may want to learn an Auto-Regressive with eXogenous input (ARX) model, to describe the mean $\mu_t$ of $Y_t$, based on lagged values of the target variable (say, one lag only), as well as lagged input features from the support agents. A first support agent owns feature $x_1$ while a second support agent owns feature $x_2$. This yields
\begin{equation} \label{eq:lrex}
    \mu_t = \underbrace{\beta_0 + \beta_1 \, y_{t-1}}_{\text{central agent}} \, + \underbrace{\beta_2 \, x_{1,t-1}}_{\text{first support agent}} \, + \underbrace{\beta_3 \, x_{2,t-1}}_{\text{second support agent}} \, .
\end{equation}
\end{example}

\begin{example}[Polynomial quantile regression of order 2] In a quantile regression problem, for a given nominal level $\tau$ (say, for instance, $\tau=0.9$), to describe the quantile $q^{(\tau)}_t$ of $Y_t$, the central agent owns feature $x_1$. In parallel, two support agents own two relevant features $x_2$ and $x_3$. Those are overall considered within the following polynomial quantile regression problem or order 2:
\begin{align} \label{eq:arxex}
    q^{(\tau)}_t  =& \quad \underbrace{\beta_0 + \beta_1 x_{1,t} + \beta_2 x_{1,t}^2}_{\text{central agent}} \, + \, \underbrace{\beta_3 x_{2,t} + \beta_4 x_{2,t}^2}_{\text{first support agent}}  \\
    &  \, + \underbrace{\beta_5 x_{3,t} + \beta_6 x_{3,t}^2 }_{\text{second support agent}}  \, + \, \underbrace{\beta_7 x_{2,t} x_{3,t} + \beta_8 x_{1,t} x_{2,t} + \beta_9 x_{1,t} x_{3,t}}_{\text{interaction terms}}  \nonumber  \, .
\end{align}
\end{example}

\subsection{Estimation problems}

For the regression problems in the above, one eventually has to estimate the parameter vector $\boldsymbol{\beta}$ based on available data. We differentiate two cases: batch and online, which are further described in the following. 

\subsubsection{Residuals and loss functions}

Eventually, based on those collected data, one aims at finding the ``best" mapping $f$ that describes the relationship between the input features and the target variable. Given a chosen regression model for $f$ (within our restricted class of regression models), this is done by minimizing a chosen loss function $l$ of the residuals $\varepsilon_t = y_t - \boldsymbol{\beta}^\top \tilde{\mathbf{x}}_t$ in expectation, to obtain the optimal set of parameters $\hat{\boldsymbol{\beta}}$, i.e.,
\begin{equation} \label{eq:regest}
    \hat{\boldsymbol{\beta}} = \argmin_{\boldsymbol{\beta}} \mathbb{E} \left[ l(\varepsilon) \right]  \, .
\end{equation}

Common loss functions include the quadratic loss $l(\varepsilon) = \varepsilon^2$ for mean regression, the absolute loss $l(\varepsilon) = |\varepsilon|$ for median regression and more generally the quantile loss $l(\varepsilon; \tau) = \varepsilon(\tau- \mathbf{1}_{\{\varepsilon \leq 0\}})$ for quantile regression. In all cases, $l$ is a negatively-oriented proper scoring rule, with minimum value at $\varepsilon=0$. It is negatively oriented since lower values are preferred (in other words, the model more accurately describes the data at hand in the sense of $l$). It is a strictly proper scoring rule since the best score value is only given to the best outcome (in principle, $\varepsilon=0$) \citep{Gneiting2007}. In the following, we will use the notation $l(\boldsymbol{\beta})$ instead, since given the explanatory and response variable data, the loss actually is a direct function of the vector of coefficients $\boldsymbol{\beta}$ only. 

The quadratic loss function readily allows for both batch and online estimation approaches, though the online case is not straightforward if considering absolute and quantile loss functions. Indeed, to use the type of gradient-based approach described hereafter, the following assumption is necessary.

\begin{assumption} \label{ass:diff}
Loss functions $l$ are twice differentiable everywhere and continuous, $l \in \mathcal{C}^2$.
\end{assumption}

Absolute and quantile loss functions do not satisfy Assumption~\ref{ass:diff}. However, one can use the smooth quantile loss introduced by \cite{Zheng2011} instead (also covering the absolute case for $\tau=0.5$). The smooth quantile loss function is defined as
\begin{equation} \label{eq:smoothquantileloss}
    l(\varepsilon;\tau,\alpha) = \tau \varepsilon + \alpha \log \left( 1 + \text{exp}\left(-\frac{\varepsilon}{\alpha}\right) \right) \, ,
\end{equation}
where $\tau$ is the nominal level of the quantile of interest, $\tau \in [0,1]$, while $\alpha \in \mathbb{R}_*^+$ is a smoothing parameter. A number of interesting properties of such loss functions, as well as relevant simulation studies, are gathered in \cite{Zheng2011}.

\subsubsection{Batch estimation}
\label{sssec:offlnest}

In the batch estimation case, the parameters of the regression model~\eqref{eq:regressionaugm} are estimated once for all based on observations gathered for times $t=1\hdots,T$. Given a choice of a regression model based on a set of features $\omega \subseteq \Omega$, we write $\boldsymbol{\beta}_{\omega}$ the vector of parameters corresponding to the potentially augmented vector of features $\tilde{\mathbf{x}}$. Given a loss function $l$, the vector of parameters can be obtained as
\begin{equation} \label{eq:batchcentral}
    \hat{\boldsymbol{\beta}}_{\omega} =  \argmin_{\boldsymbol{\beta}_{\omega}} \, L_\omega (\boldsymbol{\beta}_\omega) \, ,
\end{equation}
where $L_\omega (\boldsymbol{\beta}_\omega)$ is an in-sample estimator for $\mathbb{E} \left[ l_\omega (\boldsymbol{\beta}_\omega) \right] $, defined as
\begin{equation} \label{eq:batchcentral2}
     L_\omega (\boldsymbol{\beta}_{\omega}) = \frac{1}{T} \sum_{t=1}^T l(y_t -\boldsymbol{\beta}_\omega^\top \tilde{\mathbf{x}}_{\omega,t}) = \frac{1}{T} \sum_{t=1}^T l_{\omega,t}(\boldsymbol{\beta}_\omega) \, ,
\end{equation}
and where $\tilde{\mathbf{x}}_{\omega,t}$ is the augmented feature vector value at time $t$. We denote by $L_{\omega}^*$ the value of the loss function estimate $L_\omega$ at the estimated $\hat{\boldsymbol{\beta}}_{\omega}$, $L_{\omega}^* = L_{\omega}(\hat{\boldsymbol{\beta}}_{\omega})$. Interesting special cases then include the estimation of $\hat{\boldsymbol{\beta}}_{\omega_i}$, i.e., using the features of the central agent only with loss function value estimate $L_{\omega_i}^*$, as well as the case for which all features are considered (from both central and support agents) yielding the estimated coefficients $\hat{\boldsymbol{\beta}}_{\Omega}$ and loss function value estimate $L_{\Omega}^*$. The overall added value of employing features from support agents can then be quantified as $L_{\omega_i}^* - L_{\Omega}^*$. One may intuitively expect that all potential features $x_k \in \Omega_{-i}$ contribute to lowering the loss function estimate from $L_{\omega_i}$ to $L_{\Omega}^*$. However, such features will contribute to a varied extent, with possibly some that provide a negative contribution, i.e., in practice, they make the loss function estimate worse. It is a general problem in statistical learning and forecasting to select the right features to lower the loss function at hand.

An important property of the batch estimation problems, with model types and loss functions we consider, is described in the following proposition.

\begin{proposition} \label{prop:batchexun}
Given a convex loss function $l$ and a parametric regression model of the form of~\eqref{eq:regressionaugm}, the vector $\hat{\boldsymbol{\beta}}_{\omega}$ of optimal model parameters, as in~\eqref{eq:batchcentral}, exists and is unique. 
\end{proposition}

We do not give a formal proof of Proposition~\ref{prop:batchexun} here, as it is a straightforward and key result of convex optimization: the optimization problem in~\eqref{eq:batchcentral}, based on convex loss functions (as used in regression model estimation, e.g., quadratic, quantile, etc.), relies on a continuous and strictly convex function $L_\omega$. Hence, its solution exists and is unique.

Depending on the loss function $l$ and its in-sample estimate $L$, the estimation problem in~\eqref{eq:batchcentral} may have a closed-form solution (as for the quadratic loss case), or may require the use of numerical methods (i.e., for absolute and quantile loss functions, possibly Huber loss \citep{Huber1964} and more general convex loss functions).

\subsubsection{Online estimation}
\label{sssec:onlinest}

So far, it was assumed that the regression model parameters do not change with time. However, due to nonstationarity in the data and underlying processes, and possibly to lighten the computation burden, it may be relevant to consider that these model parameters vary in time. In that case, we also use a time index subscript for $\boldsymbol{\beta}_{\omega,t}$. The estimation of $\boldsymbol{\beta}_{\omega,t}$ in a recursive and adaptive manner is referred to as online learning. For a thorough recent coverage of approaches to online learning, the reader is referred to \cite{Orabona2020}. 

\textcolor{black}{In the online learning setup, recursivity translates to the idea that the model parameter estimates at a given time $t$ can be obtained based on the previous model parameter estimates (hence, at time $t-1$) and the new available information at time $t$. That newly available information at time $t$ typically is a function of the latest residual, i.e., the difference between the latest regression output (for time $t$, based on model parameters from time $t-1$) and the observation at time $t$. In parallel, adaptivity is linked to the use of a forgetting scheme, so that higher weight is given to the most recent information. The most usual approach is exponential forgetting, where the importance given past information decreases exponentially. It uses a forgetting factor $\lambda \in [0,1)$, with values close to 1. Past information is then weighted by $\lambda^{\delta_t}$ where $\delta_t$ denotes the age of information compared to current time $t$. Eventually, the optimization related to the estimation of model parameters at time $t$ can be be formulated as}
\begin{equation} \label{eq:regestonline}
    \hat{\boldsymbol{\beta}}_{\omega,t} = \argmin_{\boldsymbol{\beta}_{\omega,t}} \,  L_{\omega,t} (\boldsymbol{\beta}_{\omega,t}) \, ,
\end{equation}
where
\begin{equation} \label{eq:regestonline2}
L_{\omega,t} (\boldsymbol{\beta}_{\omega,t}) = \frac{1}{n_\lambda} \,  \sum_{t_i<t} \lambda^{t-t_i} \, l(y_{t_i} - \boldsymbol{\beta}_{\omega,t}^\top \tilde{\mathbf{x}}_{\omega,t_i}) = \frac{1}{n_\lambda} \,  \sum_{t_i<t} \lambda^{t-t_i} \,  l_{\omega,t_i}(\boldsymbol{\beta}_{\omega,t}) \, .
\end{equation}
In the above, $\delta_t = t-t_i$, and $n_\lambda$ is the effective window size, $n_\lambda=(1-\lambda)^{-1}$. It is a scaling parameter for the loss function estimate similar to the number of observations $T$ in the case of the batch estimator. $L_{\omega,t}$ is to be seen as a time-varying estimator of the loss function $l$ at time $t$.  

As a proxy to solving~\eqref{eq:regestonline}, one can use a fairly straightforward trick for recursive updates of all quantities involved. Given that Assumption~\ref{ass:diff} is satisfied, recursive updates can be obtained based on a Newton-Raphson step. Considering a model based on the set of features $\omega$, and with loss function $l$ (and estimator $L$), that Newton-Raphson step forms the basis for the update of model parameters $\boldsymbol{\beta}_{\omega,t-1}$ from time $t-1$ to time $t$, with
\begin{equation}\label{eq:nrstep}
    \hat{\boldsymbol{\beta}}_{\omega,t} = \hat{\boldsymbol{\beta}}_{\omega,t-1} - \frac{\nabla L_{\omega,t} (\hat{\boldsymbol{\beta}}_{\omega,t-1})}{\nabla^2 L_{\omega,t} (\hat{\boldsymbol{\beta}}_{\omega,t-1})} \, .
\end{equation}
In practice, this means that, if having the set of optimal model parameters $\hat{\boldsymbol{\beta}}_{\omega,t-1}$ at time $t-1$, one can use the above update to obtain the optimal model parameters $\hat{\boldsymbol{\beta}}_{\omega,t}$ at time $t$. Obviously, there may be a tracking error involved, which is today commonly studied in terms of regret -- see \citet{Orabona2020} for instance.

Considering both quadratic loss and smooth pinball loss functions, we have the following general results for online learning based on a Newton-Raphson step for regression models that are linear in their parameters. \textcolor{black}{In both cases, online learning based on a Newton-Raphson step requires a memory in the form of a matrix $\mathbf{M}_{\omega,t} \in \mathbb{R}^{n \times n}$, directly relating to the Hessian $\nabla^2 L_{\omega,t}$ for the loss function considered, and at time $t$.}

\begin{proposition}  \label{prop:recupdates}
Given a loss function $l$, $l \in \mathcal{C}^2$, and a regression model as in \eqref{eq:regressionaugm}, with a set $\omega$ of parameters, the Newton-Raphson step at time $t$ is given by
\begin{subequations}
\begin{eqnarray} 
\varepsilon_{\omega,t} & = & y_t - \hat{\boldsymbol{\beta}}_{\omega,t-1}^\top \tilde{\mathbf{x}}_{\omega,t}  \, ,\\
\mathbf{M}_{\omega,t} & = & \lambda \mathbf{M}_{\omega,t-1} +  \tilde{\mathbf{x}}_{\omega,t} \tilde{\mathbf{x}}_{\omega,t}^\top \, h_2(\varepsilon_{\omega,t})  \, ,\\
\hat{\boldsymbol{\beta}}_{\omega,t} & = & \hat{\boldsymbol{\beta}}_{\omega,t-1} + \mathbf{M}_{\omega,t}^{-1} \tilde{\mathbf{x}}_{\omega,t} \, h_1(\varepsilon_{\omega,t}) \, . \label{eq:updates}
\end{eqnarray}
\end{subequations}
with, if $l$ is the quadratic loss,
\begin{subequations}
\begin{eqnarray}
h_1(\varepsilon_{\omega,t}) & = & \varepsilon_{\omega,t} \, ,\\
h_2(\varepsilon_{\omega,t}) & = & 1 \, ,
\end{eqnarray}
\end{subequations}
and if $l$ is the smooth quantile loss, given the smoothing parameter $\alpha$ and nominal level $\tau$,
\begin{subequations}
\begin{eqnarray}
h_1(\varepsilon_{\omega,t}) & = & \tau +  \frac{ \alpha - \exp \left( -\frac{\varepsilon_{\omega,t}}{\alpha} \right) }{ 1 + \exp \left( -\frac{\varepsilon_{\omega,t}}{\alpha} \right) } \, , \\
h_2(\varepsilon_{\omega,t}) & = & \frac{ (1+\alpha) \exp \left( -\frac{\varepsilon_{\omega,t}}{\alpha} \right) }  { \left( 1 + \exp \left( -\frac{\varepsilon_{\omega,t}}{\alpha} \right) \right)^2 } \, .
\end{eqnarray}
\end{subequations}
\end{proposition}

There also, the proof of Proposition~\ref{prop:recupdates} is omitted, since only relying on calculating relevant derivatives and Hessian of loss functions, to be plugged in~\eqref{eq:nrstep}. Similar derivations can be performed for other types of loss functions that meet Assumption~\ref{ass:diff}, as well as for special cases of loss functions that do not meet Assumption~\ref{ass:diff}, e.g., the Huber loss. Similarly to the batch case, that approach enjoys the interesting property of existence and uniqueness of the Newton-Raphson step.

\begin{proposition} \label{prop:onlineexun}
Given a loss function $l$ that meets Assumption~\ref{ass:diff}, and a regression model as in \eqref{eq:regressionaugm}, with a set $\omega$ of parameters, the Newton-Raphson step is always feasible, while the updated vector of estimated model parameters $\hat{\boldsymbol{\beta}}_{\omega,t}$ exists and is unique.
\end{proposition}

The proof of Proposition~\eqref{prop:onlineexun} readily relies on Assumption~\ref{ass:diff}, since for loss functions $l \in \mathcal{C}_2$, both gradient $\nabla L_{\omega,t} (\hat{\boldsymbol{\beta}}_{\omega,t-1})$ and Hessian $\nabla^2 L_{\omega,t} (\hat{\boldsymbol{\beta}}_{\omega,t-1})$ are always well-defined.

\textcolor{black}{The recursive updates given in Proposition~\ref{prop:recupdates} are for a given time $t$. However, it does not tell us how such an online learning scheme should be initialized. In practice, one generally uses batch estimation with a small sample of data (say, 50 to 100 time points) to obtain initial parameter estimates for the online learning scheme. Alternatively, all parameter estimates may be initialized to 0 (or any other relevant expert guess) and the online learning scheme applied from the start. In that case though, one would need to wait for some steps (again, say, 20 to 100 time points) before to inverse a matrix in~\eqref{eq:updates}, as those may be (close to) singular.}

\subsection{Defining regression tasks}

Let us close this section related to regression by defining regression tasks, in both batch and online versions. The reason why we need to define those tasks is that these will be the tasks that central agents may post on a collaborative analytics platform, within the market frame to be described in the following section. Another type of task is finally defined for the out-of-sample regression case, when the models and estimated parameters (from either batch or online learning stage) are to be used for out-of-sample genuine forecasting.

\begin{definition}[Batch regression task]
Given the choice of regression model $f$ and loss function $l$, as well as data collected for a set of input features $x_k \in \omega \subseteq \Omega$ and a target variable $y$ over a period with $T$ time steps, a batch regression task can be represented as 
\begin{equation}
    \mathcal{F}^{\text{b}}_l: \left(\tilde{\mathbf{X}}_\omega,\mathbf{y}\right) \rightarrow \left(\hat{\boldsymbol{\beta}}_\omega, L_\omega^* \right) \, ,
\end{equation}
i.e., as a mapping from those data to a set of coefficients $\hat{\boldsymbol{\beta}}_\omega \in \mathbb{R}^n$ such that the loss function estimate is minimized (and with minimum value $L_\omega^*$).
\end{definition}

\begin{definition}[Online regression task]
At time $t$, given a regression model $f$, a loss function $l$ and a forgetting factor $\lambda$, as well as newly collected data at time $t$ for a set of input features $x_k \in \omega \subseteq \Omega$ and a target variable $y$, the online regression task relies on the following mapping
\begin{equation}
    \mathcal{F}^{\text{o}}_{l,t}: \left(\hat{\boldsymbol{\beta}}_{\omega,t-1}, L^*_{\omega,t-1}, \mathbf{M}_{\omega, t-1}, \tilde{\mathbf{x}}_{\omega,t}, y_t\right) \rightarrow \left(\hat{\boldsymbol{\beta}}_{\omega,t}, L^*_{\omega,t}, \mathbf{M}_{\omega,t} \right) \, ,
\end{equation}
where as input $\hat{\boldsymbol{\beta}}_{\omega,t-1}$ is the previous set of estimated model parameters (from time $t-1$), $L_{\omega,t-1}$ is the loss function estimate value at time $t-1$, $\mathbf{M}_{\omega, t-1} \in \mathbb{R}^{n \times n}$ is the memory of the regression task, while $\tilde{\mathbf{x}}_{\omega,t} \in \mathbb{R}^n$ and $y_t \in \mathbb{R}$ are the new data (for both input features and target variable) at time $t$. Based on those, the regression task $\mathcal{F}_{l,t}^{\text{o}}$ updates the memory to yield $\mathbf{M}_{\omega,t}$, the estimated model parameters to yield $\hat{\boldsymbol{\beta}}_{\omega,t}$, as well as the loss function estimate $L^*_{\omega,t}$.
\end{definition}

Note that the choice of regression model for $f$ and a loss function $l$ leads to a unique mapping $\mathcal{F}^b_l$ and $\mathcal{F}^o_{l,t}$ for both batch and online regression tasks, based on Propositions~\ref{prop:batchexun} and~\ref{prop:onlineexun}, respectively. In a last stage, let us define in the following the out-of-sample regression task.

\begin{definition}[Out-of-sample regression task]
At time $t$, given a choice of regression model $f$ and estimated parameters available at that time (from either batch or online regression tasks, which we write $\hat{\boldsymbol{\beta}}_{\omega,t}$), as well as data collected for a set of input features $\tilde{\mathbf{x}}_{\omega,t+h}$, the out-of-sample regression task maps those to a $h$-step ahead forecast for a characteristic $z$ of the target variable $y$, i.e., 
\begin{equation}
    \mathcal{F}^{\text{oos}}_t: \left( \tilde{\mathbf{x}}_{\omega,t+h},\hat{\boldsymbol{\beta}}_{\omega,t} \right) \rightarrow \hat{z}_{t+h|t} \, .
\end{equation}
\end{definition}

There again, the mapping exists and is unique (unless the parameters are equal to 0), since dealing with regression models that are linear in their parameters.

\section{Introducing regression markets}
\label{sec:regmarket}

\subsection{General considerations}

Emphasis is placed on a market with a single buyer and multiple sellers. This market is hosted by an analytics platform, handling both the collaborative analytics and the market components. This is in line with other works that look at data markets with some form of collaborative analytics involved as for, e.g., \citet{Agarwal2019} and \citet{Goncalves2020}.

\textcolor{black}{On this platform, a central agent $a_i$ posts a regression task (either batch or online, as defined in the above), which therefore implies a choice for a regression model $f$. This choice for $f$ is to be understood as choosing a class of potential regression models, e.g., plain linear regression or additive spline regression, based on features that may be provided.} The central agent additionally declares a willingness to pay $\phi_i$ for improving model fitting, or forecast accuracy, in the sense of a loss function $l$. The willingness to pay may be readily linked to the perceived cost of modeling and forecasting errors in some decision process, for instance if trading in electricity markets. $\phi_i$ is expressed in monetary terms (e.g., \euro{},  \textsterling{} or \$) per unit improvement in $l$ and per data point provided. If support agents were to provide relevant additional features, the loss function $l$ (or its estimate in practice) may be lowered, and the support agents remunerated based on the valuation of their features, relatively to others' features and to the overall improvement in the loss function $l$. 
\textcolor{black}{Obviously, a general problem for any statistical and machine learning setup is to select features that are valuable, here within the class of regression models considered. Otherwise, the loss function $l$ will worsen. Consequently, those features have no value to the central agent, and the support agents should not be remunerated for features that are not valuable. Within the class of regression models chosen by the central agent, the analytics platform performs the necessary feature selection, based on cross-validation for instance. Consequently, we formulate the following crucial assumption.}

\begin{assumption} \label{ass:valuable}
\textcolor{black}{Within our regression markets, given that central agents have expressed a choice for a class of regression models, the analytics platform is entrusted with the feature selection process, for instance based on cross-validation, so that only valuable features (in the sense that using them lowers the loss function $l$) are considered.}
\end{assumption}

It should be noted regression markets could endogenously perform the feature selection process, since, as will be described in the following, features that are not valuable will yield null or even negative payments. Hence, at this stage such features could be removed and the regression market run again without them. Alternatively, regression markets could rely on penalized regression problems, e.g., lasso \citep{Tibshirani1996} and elastic nets \citep{Zou2005}. This would have the advantage of endogenously selecting features, though decreasing overall benefits and potential distorting payments as the price to pay for that penalization.

\textcolor{black}{All agents involved, i.e., both central and support agents, are to be seen as opportunistic. By this, we mean that they all hope to have a gain or a payment from participation in the regression market, although no gain or payment is guaranteed. In practice, the central agent cannot know in advance whether support agents may bring valuable features and data which would improve model fitting and improve forecast accuracy (in the sense of lowering a loss function $l$). Similarly, support agents cannot know in advance whether their data and features will be selected, and what potential payment they may receive. This aspect actually is in line with other proposals for data markets with a central analytic components, e.g., \citet{Agarwal2019} and \citet{Goncalves2020}. There, the buyers place a bid and the payment to the support agents (referred to as data sellers) is readily linked to the market-clearing price. Then, if the price out of the market clearing is higher than the price offered by the buyer, the value of their input data is altered by adding a noise to it (the variance of which is proportional to the difference between the bid and actual market price). Importantly also, the market price in each trade is purely dependent on the value of the data in the previous trades, hence set before the current buyer enters the market with a specific analytics task. In the case where support agents would like to condition their participation to a minimum payment, one could also use the concept of ``reservation to sell" placed within a lasso-based regression framework, as recently proposed by \citet{Han2021}. More generally, minimum gain and payments for all agents involved could also be considered at the feature selection stage. This reservation to sell and minimum requested payment may reflect perceived privacy loss, a loss of a competitive advantage, the cost of acquiring and storing the data, etc.}

In the following, we first consider the batch setup, which allows to introduce the relevant markets concepts and its desirable properties. It is then extended to the online case, for which the data is streaming. Hence, the regression model parameters, allocation policies and payments are updated each and every time $t$, when new data becomes available. \textcolor{black}{In both cases, these markets are for an in-sample assessment of the value of the features and data of the support agents. This in-sample assessment is only a proxy for what their value may be out-of-sample, when used for genuine forecasting. In practice, there may be substantial differences between in-sample and out-of-sample estimates for a chosen loss function $l$, and this is why we consider here complementary regression markets for the in-sample and out-of-sample stages. However, it is clear that such out-of-sample forecasts can only be issued if the features and data of support agents has already been used to train relevant regression models. This is why, in our proposal for regression markets, one needs to combine an in-sample assessment of improvements in $l$ (based on the batch or online regression market) and an out-of-sample assessments of $l$ (based on an out-of-sample regression market). Our proposal for the definition of payments then rely on considerations related to quality, i.e., in-sample and out-of-sample reduction in a loss function $l$, and volume since the payment will be proportional to the quantity of data being shared by the support agents. Especially in a data streaming and online environment, this volumetric side of the payment is important to ensure that the data is continuously being shared by support agents.}

\subsection{Batch regression market}

In a batch regression market, the central agent has a willingness to pay $\phi_i$ for improving the value of the loss function $l$, for instance expressed in \euro{} per time instant (or data point) and per unit decrease in $l$. Obviously, $l$ is in practice replaced by its estimate $L$. And, the process is based on a batch of data for the time instants between times 1 and $T$. In principle, the support agents have a willingness to sell $\phi_j$ ($\forall j \neq i$), for instance expressed in \euro{} per data point shared, which may be a function of the cost of collected the data, privacy-related considerations, etc. However here, we consider that that their willingness to sell is $\phi=0$, i.e., they are happy to receive any possible payment for their features and data.

The central agent communicates the loss function $l$, regression model for $f$, length of dataset $T$ (and the actual time period it corresponds to), owns set $\omega_i$ of features, as well as willingness to pay $\phi_i$, to the analytics platform. The mapping $\mathcal{F}^b_l$ is then well-defined within that analytics platform. In parallel, interested support agents share the data for their sets $\omega_j$ of features (so, $T$ data points per feature) with the analytics platform. Within that framework, let us formally define a batch regression market.

\begin{definition}[batch regression market]
Given a regression model $f$, a loss function $l$ and a batch period with $T$ time steps, a batch regression market mechanism is a tuple ($\mathcal{R}_y$, $\mathcal{R}_x$, $\boldsymbol{\Pi}$) where $\mathcal{R}_y$ is the space of the target variable, $\mathcal{R}_y \subseteq \mathbb{R}^T$, $\mathcal{R}_x$ is the space of the input features, $\mathcal{R}_\mathbf{x} \subseteq \mathbb{R}^T$, and $\boldsymbol{\Pi}$ is the vector of payout functions $\Pi_k: \left( \{\mathbf{x}_k\}_k \in \mathcal{R}_x^{K}, \mathbf{y} \in \mathcal{R}_\mathbf{y} \right) \rightarrow \pi_k \in \mathbb{R}^+$.
\end{definition}

Based on all features provided, and based on the mapping $\mathcal{F}^b_l$, the analytics platform deduces the overall improvement in the loss function estimate $L$ as $L^*_{\omega_i}-L^*_{\Omega}$. This yields the payment $\pi_i$ of the central agent 
\begin{equation} \label{eq:capayment}
    \pi_i = T \, (L^*_{\omega_i}-L^*_{\Omega}) \, \phi_i \, ,
\end{equation}
which is a direct function of the quantity of data, improvement in loss function $l$ (as estimated over the data used for estimated the model parameters) and the willingness to pay of the central agent. \textcolor{black}{As mentioned previously, the payment has a volumetric component, since one buys a quantity of $T$ data points at once, and a quality component, since the payment is conditioned by the decrease in the loss function estimate by using the features and data of the support agents.}

In parallel, the batch regression market relies on allocation policies $\psi_k (l)$ to define the payment for any feature $x_k$ of the support agents, $x_k \in \Omega_{-i}$. We write $\psi_k (l)$ the allocation policy value for feature $x_k$ for the loss function $l$, corresponding to its marginal contribution to the overall decrease of the loss function $S$. We therefore intuitively expect the following desirable properties for allocation policies.

\begin{property} \label{prop:allocationpolicies}
Allocation policies $\psi_k(l)$ are such that
\begin{itemize}
    \item[(i)] $\psi_k (l) \in [0,1] \, ,  \, \,  \forall k$  
    \item[(ii)] $\sum_k \psi_k (l)=1$  
\end{itemize}
\end{property}

Those desirable properties for allocation policies are crucial for some of resulting inherent properties of the regression markets to be introduced and discussed later on. Eventually, the payment for feature $x_k$ is
\begin{equation} \label{eq:paymentbatch}
    \pi_k =  T \, (L^*_{\omega_i}-L^*_{\Omega}) \, \phi_i \,  \psi_k (l) \, , \quad \forall k \in \Omega_{-i} \, ,
\end{equation}
so that, overall, the payment to agent $j$ is 
\begin{equation} \label{eq:paymentbatchagentj}
    \pi (a_j) = T \,  (L^*_{\omega_i}-L^*_{\Omega}) \, \phi_i \,  \left( \sum_{k|x_k \in \omega_j} \psi_k (l) \right)\, , \quad \forall a_j \in \mathcal{A}_{-i}. 
\end{equation}
The payment is both volumetric, since the quantity of data $T$ is accounted for and linearly influences the payment, as well as quality driven. On that last point, it is a function of the overall improvement in $l$ by considering the support agents' features (i.e., $L^*_{\omega_i}-L^*_{\Omega}$), and the marginal contribution of each and every feature $x_k$ to that improvement (through the allocation policy $\psi_k(l)$).

The key question is then about how to value each and every feature from the support agents within the regression task at hand (and hence, for the central agent). This is the aim of the allocation policies $\psi_k$ in~\eqref{eq:paymentbatch} and~\eqref{eq:paymentbatchagentj}. 

In the simplest case where the input features are independent, the regression model separable and linear, and a quadratic loss is considered, one may actually consider the coefficient of determination as a basis for determining the $\psi_k$'s. We refer to this approach as a ``leave-one-out" policy. 

\begin{definition}[leave-one-out allocation policy]
For any feature $x_k \in \Omega_{-i}$, and loss function $l$, the leave-one-out allocation policy $\psi^{\text{loo}}_k (l)$ can be estimated as
\begin{equation} \label{eq:loo1}
    \psi^{\text{loo}}_k (l) = \frac{L^*_{\Omega \setminus \{x_k\}} - L^*_{\Omega}}{L^*_{\omega_i} - L^*_{\Omega}}, \quad \text{or} \quad \psi^{\text{loo}}_k (l) = \frac{L^*_{\omega_i} - L^*_{\omega_i \cup \{x_k\}}}{L^*_{\omega_i} - L^*_{\Omega}} \, .
\end{equation}
\end{definition}

In the above, both estimators are scaled by the loss estimate improvement when going from the central agents features only ($\omega_i$) to the whole set of features $\Omega$. The difference between the two estimators is in the numerator. In the first case, $L^*_{\Omega \setminus \{x_k\} - L^*_{\Omega}}$ is for the decrease in the loss estimate when going from the full set of features minus $x_k$ to the full set of features. And, in the second case, $L^*_{\omega_i} - L^*_{\omega_i \cup \{x_k\}}$ is for the decrease in the loss estimate when going from the set of features of the central agent only, to that set plus $x_k$. This leave-one-out policy may be seen as a simple case of a Vickrey-Clarke-Grove (VCG) mechanism, and for instance considered by \citet{Agarwal2019} and \citet{Rasouli2021}.

For the special case where $l$ is a quadratic loss function, one can take a variance-decomposition point of view to observe that
\begin{equation} \label{eq:loo2}
    \psi^{\text{loo}}_k (l) = \frac{\beta_k^2 \,  \text{Var}[X_k]}{\sum_{j \in \Omega_{-i}} \beta_j^2 \, \text{Var}[X_j]},
\end{equation}
with $\text{Var}[.]$ the variance operator. Hence, $\psi^{\text{loo}}_k (l)$ readily translates to the share of the variance in the target variable explained by the feature $x_k$. Consequently, both estimators are equivalent and one readily verifies that allocation policies fulfil Property \ref{prop:allocationpolicies}.

Strictly speaking, the leave-one-out allocation policies do not meet the desirable properties expressed in Property~\ref{prop:allocationpolicies}, unless Assumption~\ref{ass:valuable} is respected. It may not even be appropriate in the case where the features are not independent, when the regression model is non-separable and nonlinear, and if the loss function is not quadratic. In that more general case, a Shapley-based approach can be used instead. Shapley values and related allocation are well-known concepts in cooperative game theory with many desirable properties, while essentially providing a fair compensation for an agent's contribution to collective value creation. For a compact introduction, the reader is referred to \citet{Winter2002}, while the application of Shapley value for data valuation is covered by \citet{Ghorbani2019}.

Allocation values are consequently defined by the marginal value of the various features in a Shapley sense, hence yielding the Shapley allocation policy.

\begin{definition}[Shapley allocation policy]
For any feature $x_k \in \Omega_{-i}$, and loss function $l$, the (original) Shapley allocation policy $\psi^{\text{sh}}_k (l)$ is given by
\begin{equation} \label{eq:shapley}
    \psi^{\text{sh}}_k (l) = \frac{1}{L^*_{\omega_i} - L^*_{\Omega}} \sum_{\omega \subseteq \Omega_{-i} \setminus \{x_k\}} \frac{|\omega|! (|\Omega_{-i}|-|\omega|-1)!}{|\Omega_{-i}|!} \left( L^*_{\omega_i \cup \omega} - L^*_{\omega_i \cup \omega  \cup \{x_k\}}\right) \, .
\end{equation}
\end{definition}
In the case where features are independent, considering a linear regression and a quadratic loss function, one has $\psi^{\text{sh}}_k (l) = \psi^{\text{loo}}_k (l)$. Even in the linear case and with quadratic loss, if features are not independent, spurious allocation may be obtained when employing the leave-one-out strategy, as hinted by \citet{Agarwal2019}. For instance, consider two features $x_k$ and $x_{k'}$ being correlated perfectly, the marginal value of each feature as given by $\psi^{\text{loo}}_k (l)$ and $\psi^{\text{loo}}_{k'} (l)$ would be 0 if using the first estimator in~\eqref{eq:loo1}. In contrast if using the second estimator in~\eqref{eq:loo1}, $\psi^{\text{loo}}_k (l)$ and $\psi^{\text{loo}}_{k'} (l)$ would correctly reveal their marginal value, but one would eventually have $\sum_j \psi^{\text{loo}}_k (l)>1$ (since considering twice the same marginal feature value in the overall picture), which does not respect the basic definition such that allocations should sum to 1. The reason why we introduce here those two types of allocations is that in practice, various allocations could be used alternatively, as long as allocation policies are positive and sum to 1. Despite the fact Shapley allocations should be seen as the most relevant one, these are notoriously computationally heavy to calculate as the number of features $n$ increases. This is general problem known and addressed by the computer science and algorithmic game theory communities, see e.g., \cite{Jia2019} for a recent example also related to data markets.

A more important issue with the Shapley allocation policy is that it may violate one of the desirable basic properties of allocation policies, i.e., such that $\phi_k \in [0,1]$. This is since, as indicated in Section~\ref{sssec:offlnest}, certain features may actually make the loss function estimate worse when they provide no (or very little, compared to the batch sample size) valuable information. For those features, readily using a Shapley allocation policy would yield negative values for $\phi_k$. This problem was for instance recently identified and discussed by \citet{Liu2020}, who then proposed to use zero-Shapley and absolute-Shapley values instead.

\begin{definition}[zero-Shapley and absolute-Shapley allocation policies]
For any feature $x_k \in \Omega_{-i}$, and loss function $l$, the zero-Shapley allocation policy $\psi^{\text{sh}}_k (l)$ is given by
\begin{equation} \label{eq:shapley}
    \psi^{\text{sh}}_k (l) = \frac{1}{L^*_{\omega_i} - L^*_{\Omega}} \sum_{\omega \subseteq \Omega_{-i} \setminus \{x_k\}} \frac{|\omega|! (|\Omega_{-i}|-|\omega|-1)!}{|\Omega_{-i}|!} \, \max \left\{ 0, L^*_{\omega_i \cup \omega} - L^*_{\omega_i \cup \omega  \cup \{x_k\}} \right\} \, ,
\end{equation}
while the absolute-Shapley allocation policy $\psi^{\text{sh}}_k (l)$ is defined as
\begin{equation} \label{eq:shapley}
    \psi^{\text{sh}}_k (l) = \frac{1}{L^*_{\omega_i} - L^*_{\Omega}} \sum_{\omega \subseteq \Omega_{-i} \setminus \{x_k\}} \frac{|\omega|! (|\Omega_{-i}|-|\omega|-1)!}{|\Omega_{-i}|!} \left| L^*_{\omega_i \cup \omega} - L^*_{\omega_i \cup \omega  \cup \{x_k\}} \right| \, .
\end{equation}
\end{definition}

It is unclear today what approach to correcting Shapley allocation policies may be more appropriate when looking at data important and valuating in the context of regression markets. At least, both definitions ensure that the resulting allocation policies are positive -- they may not sum to 1 though. In our case, by using Assumption~\ref{ass:valuable}, the original Shapley allocation policy can be readily employed, while meeting Property~\ref{prop:allocationpolicies}. The zero-Shapley and absolute-Shapley allocation policies may be useful instead for the out-of-sample regression markets, where the inherent value of the features and data provided by the support agents may not necessarily be positive.

Finally, let us compile here the important properties of the batch regression market mechanism introduced in the above, which we look at in a way that is fairly similar to the case of wagering markets as in \citet{Lambert2008} as well as data markets as in \cite{Agarwal2019}.

\begin{theorem} \label{theo:batchmarket}
Batch regression markets, using the proposed regression framework and payout functions based on (original) Shapley allocation policies, yield the following desirable properties:
\begin{enumerate}
\item[(i)] {\bf Budget balance} -- the sum of revenues is equal to the sum of payments
\item[(ii)] {\bf Symmetry} -- the market outcomes are independent of the labelling of the support agents
\item[(iii)] {\bf Truthfulness} -- support agents only receive their maximum potential revenues when reporting their true feature data
\item[(iv)] {\bf Individual rationality} -- the revenue of the support agents is at least 0
\item[(v)] {\bf Zero-element} -- a support agent that does not provide any feature, or provide a feature that has no value (in terms of improving the loss estimate $L_\omega$), gets a revenue of 0
\textcolor{black}{\item[(vi)] {\bf Linearity} -- for any two set of features $\omega$ and $\omega'$, the revenue obtained by sharing $\omega \cup \omega'$ is equal to the sum of the revenues if having shared $\omega$ and $\omega'$ separately}
\end{enumerate}
\end{theorem}

Note that \citet{Lambert2008} also mentions sybilproofness, normality and monotonicity, which are not seen as relevant here. Those properties may be investigated in the future though. \textcolor{black}{Linearity is an interesting property which ensures that support agents will not be strategic in packaging their features since, whatever the way they submit features to the regression market (individually or as a bundle), the overall revenue obtained will be the same. Note also that, similarly to the case of \cite{Agarwal2019}, the batch regression markets inherit the additive property from the additivity axiom defining Shapley values.} The proofs are gathered in Appendix~\ref{sec:proofoftheorem1}. Truthfulness can only be ensured up to sampling uncertainty since, as discussed in the proof, it would strictly hold if having access to the actual loss $l$ -- in practice, however, only an in-sample estimate is available. For the case of using leave-one-out allocation policies instead, the same properties are obtained for plain linear regression models, a quadratic loss, and independent features. This relies on the law of total variance, of which the variance decomposition of~\eqref{eq:loo1} is an example consequence. Truthfulness may not be verified in the more general case, though the other properties will hold. \textcolor{black}{These properties are obviously interesting -- still, they may not prevent some potential challenges with data duplication. For strategic behavior such as data replication, we have to use other payoff allocation mechanisms, such as the Shapley Approximation proposed by \cite{Agarwal2019}, at the cost of loosing budget balance.}

Finally, it should be noted that such a setup for batch regression market may be readily extended to the case of batch learning based on sliding windows, since payments would only be due for the new data points being used.

\subsection{Online regression market}

To adapt to the fact data is naturally streaming, and also that the analytics approaches may require to continuously learn from data in an online environment, we propose here an online version of the regression market introduced in the above. The base considerations are the same. The central agent has a willingness to pay $\phi_i$ for improving the value of the loss function $l$ (in \euro{} per time instant and per unit decrease in $l$). This agent communicates the loss function $l$, regression model for $f$, her own set $\omega_i$ of features, as well willingness to pay $\phi_i$, to the analytics platform. Most likely, the central agent also needs to inform about the duration over which the process will be re-iterated, as it may not make sense to only try and learn at a single instant. On the other side, interested support agents $a_j$ share the data for their sets $\omega_j$ of features with the analytics platform, by delivering a new set of feature data at each and every time $t$, as time passes. At each time $t$, the mapping $\mathcal{F}^o_{l,t}$ is well-defined within that analytics platform. Within that framework, let us formally define an online regression market.

\begin{definition}[online regression market]
Given a regression model $f$, a loss function $l$ and a given time $t$, an online regression market mechanism is a tuple ($\mathcal{R}_y$, $\mathcal{R}_x$, $\boldsymbol{\Pi}$) where $\mathcal{R}_y$ is the space of the target variable, $\mathcal{R}_y \subseteq \mathbb{R}$, $\mathcal{R}_x$ is the space of the input features, $\mathcal{R}_\mathbf{x} \subseteq \mathbb{R}$, and $\boldsymbol{\Pi}$ is the vector of payout functions $\Pi_k: \left( \{\mathbf{x}_k\}_k \in \mathcal{R}_x^{K}, \mathbf{y} \in \mathcal{R}_\mathbf{y} \right) \rightarrow \pi_{k,t} \in \mathbb{R}^+$.
\end{definition}

\textcolor{black}{In the batch regression case, one has a single estimate of the loss $l$ over the batch period (with $T$ data points), hence allowing to define a payment (for example for a given feature $k$ in \eqref{eq:paymentbatch}) that combines the contribution to the loss improvement and the volume of data. In an online regression case, however, the loss estimator varies with time. It is hence not possible to define a single payment over a period with $T$ data points based on a single loss function value. Instead, one needs to track the loss estimates through time, and use the loss estimate at time $t$ to value the data points provided at that time. In a way, in the batch case, one could also consider that the payment $\pi_{k,t}$ to a support agent for data point at time $t$ for feature $k$ is
\begin{equation}
    \pi_{k,t} = (L^*_{\omega_i}-L^*_{\Omega}) \phi_i \psi_k (l) \, ,
\end{equation}
while the overall payment of the central agent at a given time $t$ is
\begin{equation}
    \pi_{i,t} = (L^*_{\omega_i}-L^*_{\Omega}) \phi_i \, .
\end{equation}
These payments are the same for all $t$, and by summing up over time ($t=1,\hdots,T$), one retrieves the payments defined in~\eqref{eq:capayment} and~\eqref{eq:paymentbatch}. Now, we will extend this idea of having payments at each and every time $t$ to the case of time-varying loss estimates.}

In an online learning environment, it is the loss estimators $L_{\omega,t}$ that vary with time. They will impact the allocation policies and make them time-varying too. By first observing that~\eqref{eq:regestonline2} can be decomposed as
\begin{equation} \label{eq:regestonlinedecomp}
L^*_{\omega,t} (\boldsymbol{\beta}_\omega) = \lambda \, L^*_{\omega,t-1} (\boldsymbol{\beta}_\omega) + \frac{1}{n_\lambda} l_{\omega,t}(\boldsymbol{\beta}_\omega) \, ,
\end{equation}
loss function estimates can be readily updated at each and every time $t$. Consequently, at a given time $t$, the payment $\pi_{i,t}$ of the central agent is
\begin{equation} \label{eq:capaymentonline}
    \pi_{i,t} = (L^*_{\omega_i,t}-L^*_{\Omega,t}) \, \phi_i \, .
\end{equation}
Compared to the batch case in~\eqref{eq:capayment}, $T$ has disappeared since the payment is for a single time instant, while the loss estimates are specific to time $t$. This represents a time-varying generalization of the payment for the batch case.

To obtain the payments to the support agents, the only aspect missing is to determine the allocation policies. In line with the online estimation in Section~\ref{sssec:onlinest}, which is recursive and time-adaptive, it would be ideal to have a recursive and simple approach to update allocation policies.

\begin{proposition}
At any given time $t$, both leave-one-out and (original) Shapley allocation policies can be updated in a recursive fashion, with
\begin{equation}
    \psi_{k,t} (l)  = \lambda \psi_{k,t-1} (l) + (1-\lambda) \psi_{k} (l_t) \, .
\end{equation}
\end{proposition}

This means that, for a given feature $x_k$ and both types of allocation policies, the allocation at time $t$ can be obtained based on the previous allocation at time $t-1$ and on the allocation specific to the loss $l(y_t -\boldsymbol{\beta}_\omega^\top \tilde{\mathbf{x}}_t)$ for the new residual at time $t$. Consequently, a payment $\pi_{k,t}$ (for feature $x_k$) is made at each and every time step $t$ based on the time-varying loss function estimates and allocation policies. This yields
\begin{equation} \label{eq:paymentonline-timestep-exp}
    \pi_{k,t} =  (L^*_{\omega_i,t} - L^*_{\Omega,t}) \, \phi_i \psi_{k,t} (l) \, .
\end{equation}

\textcolor{black}{Similar to the case of the model parameter estimates in online learning, a legitimate question is about how to initialize payments. Since the allocation policies and payments are readily obtained from the loss estimates (and hence model parameter estimates), the approach to model parameter initialization will drive the initialization of the payments.}

Finally, online regression markets have the same properties as the batch ones.

\begin{corollary}
Online regression markets, using the proposed regression framework and payout functions based on Shapley allocation policies, yield the properties of (i) budget balance, (ii) symmetry, (iii) truthfulness, (iv) individual rationality, (v) zero-element, and (vi) linearity.
\end{corollary}

The proof for that corollary is omitted since similar to that for Theorem~\ref{theo:batchmarket}.

\subsection{Extension to forecasting and out-of-sample loss function assessment}

Both of the above markets, in batch and online versions, relate to a learning problem and the in-sample assessment of a loss function $l$. In many practical cases, however, such models are then to be used out of sample, for forecasting purposes for instance. There may hence be a discrepancy between the in-sample loss estimate and the out-of-sample one. This is while, if forecasts are to be used as a basis for decision making, the actual perceived cost induced by the deviation between forecast and realization is represented by the out-of-sample loss, not the in-sample one.

Consequently, besides the batch and online regression markets that relate to the learning task, those should be complemented by out-of-sample payments. One can here make a direct comparison with the case of electricity markets, where one usually first has a forward (e.g., day-ahead) mechanism leading to resource allocation and payments, and then a balancing mechanism to update and correct the outcomes from the forward mechanism. In the present case, the learning process is first necessary to fit a regression model and assess the in-sample value of the features of support agents. Then, out of sample, the input features of those agents are used for genuine forecasting, and payments are to be based on the contribution to a decrease in the loss function $l$ and its out-of-sample estimate. Let us formally define the out-of-sample regression manner in the following.

\begin{definition}[out-of-sample regression market]
Given a regression model $f$ and its parameters estimated through either batch or online regression markets, a loss function $l$ and a out-of-sample period with $\left| \mathcal{T}^o \right|$ time steps, an out-of-sample regression market mechanism is a tuple ($\mathcal{R}_y$, $\mathcal{R}_x$, $\boldsymbol{\Pi}$) where $\mathcal{R}_y$ is the space of the target variable, $\mathcal{R}_y \subseteq \mathbb{R}^T$, $\mathcal{R}_x$ is the space of the input features, $\mathcal{R}_\mathbf{x} \subseteq \mathbb{R}^T$, and $\boldsymbol{\Pi}$ is the vector of payout functions $\Pi_k: \left( \{\mathbf{x}_k\}_k \in \mathcal{R}_x^{K}, \mathbf{y} \in \mathcal{R}_\mathbf{y} \right) \rightarrow \pi_k \in \mathbb{R}^+$.
\end{definition}

Consider being at a time $t$, having to use some of the regression models trained based on a batch of past data, or online. The estimated parameters are here denoted by $\hat{\boldsymbol{\beta}}_{\omega,t}$ to indicate that they are those available at that time. In the batch case, these might be older since estimated once for all on older data, unless a sliding window approach is used. In the online case instead, those may be the most recent parameters available based on the latest updated at time $t$. That model is used to issue a forecast $y_{t+h|t}$ for lead time $t+h$ for the target variable of interest, or possibly a nowcast (i.e., with $h=0$) in the case $y$ is not observed in real-time. We write $\mathcal{T}^{\text{o}}$ the set of time instants over which forecasts are being issued. The out-of-sample loss estimate over $\mathcal{T}^{\text{o}}$ is
\begin{equation} \label{eq:oosloss}
    L^{\text{o}}_{\omega} (\hat{\boldsymbol{\beta}}_{\omega,t}) = \frac{1}{|\mathcal{T}^{\text{o}}|} \sum_{t \in \mathcal{T}^{\text{o}}} l \left( y_{t+h} - \hat{\boldsymbol{\beta}}_{\omega,t}^\top \tilde{\mathbf{x}}_{t+h} \right) \, .
\end{equation}
Such an estimator is separable in time, i.e.,
\begin{equation}
L^{\text{o}}_{\omega} = \sum_{t \in \mathcal{T}^{\text{o}}} l_{\omega,t} \left( \hat{\boldsymbol{\beta}}_{\omega,t} \right)  \, , \quad \text{where} \quad 
l_{\omega,t} = \frac{1}{|\mathcal{T}^{\text{o}}|} l \left( y_t - \hat{\boldsymbol{\beta}}_{\omega,t}^\top \tilde{\mathbf{x}}_t \right) \, .
\end{equation}

Again, considering the linearity property of both leave-one-out and Shapley allocations, this translates to having over the out-of-sample period
\begin{equation}
\psi_k (l) = \sum_{t \in \mathcal{T}^{\text{o}}} \psi_{k,t} (l) \, ,
\end{equation}
where $\psi_{k,t} (l)$ is an allocation based on the evaluation of the loss function $l$ at time $t$ only. Such time-dependent allocation are then directly linked to the idea of using Shapley additive explanation \citep{Lundberg2017} for interpretability purposes. However here, such allocations aim at defining the contribution of the various features to the loss for a given forecast at time $t$. Eventually, the payment for feature $x_k$ at time $t$ (and linked to the forecast for time $t+h$) is
\begin{equation} \label{eq:paymentoos}
    \pi_{k,t} =  (l_{\omega_i,t} - l_{\Omega,t}) \, \phi_i \psi_{k,t} (l) \, .
\end{equation}
Those payments can be summed over the out-of-sample period $\mathcal{T}^{\text{o}}$, i.e.,
\begin{equation} \label{eq:paymentbatch-timestep}
    \pi_k = \sum_{t=1}^T \pi_{k,t} \, .
\end{equation}
On the central agent side, the payment at each time instant is
\begin{equation} \label{eq:paymentoos}
    \pi_{i,t} =  (l_{\omega_i,t} - l_{\Omega,t}) \, \phi_i \, ,
\end{equation}
to them be summed over the period $\mathcal{T}^\text{o}$.

Finally, based on those concepts, the out-of-sample regression markets enjoy the same desirable properties as the batch and online regression markets.

\begin{corollary}
Out-of-sample regression markets, using the proposed regression framework and payout functions based on Shapley allocation policies, yield the properties of (i) budget balance, (ii) symmetry, (iii) truthfulness, (iv) individual rationality, (v) zero-element, and (vi) linearity.
\end{corollary}

The proof for that corollary is omitted since similar to that for Theorem~\ref{theo:batchmarket}.

\section{Illustrative examples based on simulation studies}
\label{sec:simulation}

To illustrate the various regression markets, we first concentrate on a number of examples and related simulation studies. Obviously, these are simplified versions of what would be done with real-world applications, since for instance, the models of the central agent are well specified. In parallel, we focus on the batch and online regression markets only, since the use of out-of-sample markets will be more interesting and relevant when focusing on a forecasting application with real data later on.

\subsection{Batch regression market case}

In order to underline the broad applicability of the presented regression market approach, emphasis is placed on three alternative cases: plain linear regression with a quadratic loss, polynomial regression with a quadratic loss, and an autoregression with exogenous input with quantile loss.

\subsubsection{Case 1: Plain linear regression and quadratic loss}

Firstly, emphasis is placed on the simplest case of a plain linear regression problem, for which the central agent $a_1$ focuses on the mean $z$ of a target variable $Y$, while owning feature $x_1$. A quadratic loss function $l$ is used. The willingness to pay of $a_1$ is $\phi_1=0.1$\euro{} per time instant and per unit improvement in $l$. In parallel, two support agents $a_2$ and $a_3$ own relevant features $x_2$ (for $a_2$), $x_3$ and $x_4$ (for $a_3$). The regression chosen by the central agent (which is well specified in view of the true data generation process) and posted on the analytics platform relies on a model of the form
\begin{equation} \label{eq:lrexs}
    Y_t = \beta_0 + \beta_1 x_{1,t} + \beta_2 x_{2,t} + \beta_3 x_{3,t} + \beta_4 x_{4,t} +  \varepsilon_t \, , 
\end{equation}
where $\varepsilon_t$ is a realization of a white noise process, centred on 0 and with finite variance.

Let us for instance consider a case where the true parameter values are $ \boldsymbol{\beta}^\top = [ 0.1 \, \, -\!0.3 \, \, 0.5 \, \, -\!0.9 \, \, 0.2 ]$. For all features, the values of the input features are sampled from a Gaussian distribution, $ x_{j,t} \sim \mathcal{N}(0,\sigma_j^2)$, with $\sigma_j=1, \, \forall j$. In addition, $\varepsilon_t \sim \mathcal{N}(0,\sigma_\varepsilon^2), \, \forall t$, with $\sigma_\varepsilon = 0.3$.

We simulate that process for $T=$ 10 000 time steps and learn the model parameters $\boldsymbol{\beta}$ based on that period. The in-sample loss function estimates considering the central agent features only (so, an intercept and $x_1$), and then with features from the additional support agents ($x_2$, $x_3$ and $x_4$) are gathered in Table~\ref{tab:lrS}. For this specific run and example, the overall value of the support agents is (1.191-0.087) = 1.104.

\begin{table}[!ht]
    \caption{In-sample loss estimates with and without the support agents}
    \label{tab:lrS}
    \vspace{2mm}
    \centering
    \begin{tabular}{|c|c c|}
    \hline
        Agents & $\{a_1\}$ & $\{a_1,a_2,a_3\}$ \\
        In-sample loss estimate & 1.191 &  0.087\\
        \hline
    \end{tabular}
\end{table}

Since in this simple setup, we use linear regression, have independent input features and a quadratic loss function, both leave-one-out and Shapley allocation policies are equivalent. Those are gathered in Table~\ref{tab:lrfv}. This table also gathers the payments received by agents $a_2$ and $a_3$ for their features. The values for both allocation policies are the same, up to some rounding. The overall payment from the central to the support agents is of 1104\euro{} (i.e., $1.104 \times 0.1 \times$ 10 000).  

\begin{table}[!ht]
    \caption{Leave-one-out and Shapley allocation policies, on both per-feature and per-agent basis, as well as payments to the support agents.}
    \label{tab:lrfv}
    \vspace{2mm}
    \centering
    \begin{tabular}{|c|c c c|c c|}
    \hline
        Feature/Agent & $x_2$ & $x_3$ & $x_4$ & $a_2$ & $a_3$\\
        \hline
        $\psi^{\text{loo}}_k$ [\%] & 22.7 & 73.4 & 3.9 & 22.7 & 77.3\\
        $\psi^{\text{sh}}_k$ [\%]  & 22.7 & 73.4 & 3.9 & 22.7 & 77.3\\
        \hline
        $\pi^{\text{loo}}_k$ [\euro{}] & 250.7 & 810 & 43.3 & 250.7 & 853.3\\
        $\pi^{\text{sh}}_k$ [\euro{}]  & 250.7 & 810 & 43.3 & 250.7 & 853.3\\
        \hline
    \end{tabular}
\end{table}

Note that this is the only case where leave-one-out allocation policies are used, since this will not make sense for the other case studies which are more advanced, e.g., with non-separable and nonlinear regression models, and/or loss functions that are not quadratic.

\subsubsection{Case 2: Polynomial regression (order 2) and quadratic loss}

We generalize here to a polynomial regression of order 2, with the same number of agents, a quadratic loss and the same willingness to pay ($\phi_1=0.1$\euro{} per time instant and per unit improvement in $l$). The central agent $a_1$ focuses on a target variable $y$, while owning feature $x_1$. The two support agents $a_2$ and $a_3$ own relevant features $x_2$ (for $a_2$) and $x_3$ (for $a_3$). The regression chosen by the central agent and posted on the analytics platform relies on the following model:
\begin{align} \label{eq:lrexs2}
    Y_t = &  \quad \beta_0 + \beta_1 x_{1,t} + \beta_2 x_{2,t} + \beta_3 x_{3,t} + \beta_4 x_{1,t}^2+ \beta_5 x_{2,t}^2+ \beta_6 x_{3,t}^2 \\
    & + \beta_7 x_{1,t} x_{2,t} + \beta_8 x_{1,t} x_{3,t} + \beta_9 x_{2,t} x_{3,t} +  \varepsilon_t \, ,  \nonumber
\end{align}
where $\varepsilon_t$ is a realization of a white noise process, centred on 0 and with finite variance. It is well-specified and hence corresponds to the true data generation process. The true parameter values are $ \boldsymbol{\beta}^\top = [ 0.2 \, \, -\!0.4 \, \, 0.6 \, \, 0.3 \, \, 0 \, \, 0.1 \, \, 0 \, \, 0 \, \, -\!0.4 \, \, 0]$. For all features, the values of the input features are sampled from a Gaussian distribution, $ x_{j,t} \sim \mathcal{N}(0,\sigma_j^2)$, with $\sigma_j=1, \, \forall j$. In addition, $\varepsilon_t \sim \mathcal{N}(0,\sigma_\varepsilon^2), \, \forall t$, with $\sigma_\varepsilon = 0.3$.

\textcolor{black}{The process is simulated over $T=$ 10 000 time steps to estimate the regression model parameters, as well as to compute Shapley allocation policies and payments. Following Assumption~\ref{ass:valuable}, feature selection is performed and only the relevant terms in the polynomial regression (i.e., those with non-zero parameters, and contributing to lowering loss function estimates) are retained. It should be noted that there is additional subtlety in the case of interaction terms, making that one should obtain Shapley allocation policies (and payments) at the feature level (so, here, $x_1$, $x_2$ and $x_3$), and not based on individual components in the regression model. This is since these terms in the regression model may come as a bundle. For instance, if starting with a coalition with $x_3$ only, and aiming to assess the value from adding $x_2$ to that coalition, it brings 2 additional terms ($x_2$ and $x_2^2$) at once. And, if starting from a coalition with $x_1$ only, and aiming to assess the value from adding $x_3$ to that coalition, it adds 2 additional terms $x_3$ and $x_1 \, x_3$ too. However, it also means that one should look at it the other way around, and recognize that $x_1$ contributes to the value brought by the term $x_1 \, x_3$.}

\textcolor{black}{Eventually here, the in-sample loss function estimates based on the central agent features only (so, the intercept and $x_1$) is of 0.72. With features from the additional support agents, it decreases to 0.09. The value of support agents is then of 0.63 in terms of the reduction of the loss function, though part of it also comes from the central agent's feature $x_1$ through the interaction term. The Shapley allocation policy values for the various features are gathered in Table~\ref{tab:lrfv2}, as well as related payments. The contribution of feature $x_2$ comes from both $x_2$ and $x_2^2$, while that of feature $x_3$ comes from the $x_3$ and $x_1 \,x_3$ terms. However, the sum of the Shapley allocation policies is not of 1 (it is of 0.65), since part of the overall improvement comes from the central agent's feature $x_1$ through the interaction term $x_1 \,x_3$. The overall payment from the central agent is of 520.42\euro{}. The payment would have been of 630\euro{} if all improvements came from the feature and data of the support agents alone (hence, no interaction term).}

\begin{table}[!ht]
    \caption{Shapley allocation policies, on both per-feature and per-agent basis, as well as payments to the support agents.}
    \label{tab:lrfv2}
    \vspace{2mm}
    \centering
    \begin{tabular}{|c|c c|}
    \hline
        Feature (/Agent) & $x_2$ (/$a_2$) & $x_3$ (/$a_3$)\\
        \hline
        $\psi_k$ [\%]  & 44.36 & 20.59 \\
        $\pi_k$ [\euro{}]  & 355.44  & 164.98  \\
        \hline
    \end{tabular}
\end{table}

\subsubsection{Case 3: Quantile regression based on an ARX model}

In the third case, the central agent wants to learn an Auto-Regressive with eXogenous input (ARX) model with a quantile loss function (with nominal level $\tau$), based on lagged values of the target variable (say, one lag only), as well as lagged input features from the support agents. The setup with agents and features is the same as for case 1. The willingness to pay is of $\phi_1$=1\euro{} per time instant and per unit improvement in the quantile loss function. The underlying model for the regression reads
\begin{equation} \label{eq:qrexs}
    Y_t = \beta_0 + \beta_1 y_{t-1} + \beta_2 x_{2,t-1} + \beta_3 x_{3,t-1} + \beta_4 x_{4,t-1} +  \varepsilon_t \, , 
\end{equation}
where $\varepsilon_t$ is a realization of a white noise process, centred on 0 and with finite variance.

The central agent is interested in 2 quantiles, with nominal levels 0.1 and 0.75, hence requiring 2 batch regression tasks models in parallel. Support agent features are sampled similarly as in case 1 (from a standard Normal), and the characteristics of the noise term are also the same. The true parameter values are $ \boldsymbol{\beta}^\top = [ 0.1 \, \, 0.92 \, \, -\!0.5 \, \, 0.3 \, \, -\!0.1]$.

We simulate that process for 10 000 time steps. The quantile loss estimates based on the central agent features only are of 0.086 and 0.152 for the 2 nominal levels of 0.1 and 0.75. In parallel, when using the support agent features, these decrease to 0.052 and 0.096, respectively. The improvements are hence of 0.034 and 0.056 for those 2 nominal levels. The Shapley allocation policy values and payments to the support agents are gathered in Table~\ref{tab:arxqrbatch}.

\begin{table}[!ht]
    \caption{Shapley allocation policies, on both per-feature and per-agent basis, as well as payments to the support agents.}
    \label{tab:arxqrbatch}
    \vspace{2mm}
    \centering
    \begin{tabular}{|c||c|c c c|c c|}
    \hline
        $\tau$ &Feature/Agent: & $x_2$ & $x_3$ & $x_4$ & $a_2$ & $a_3$\\
        \hline
        \multirow{ 2}{*}{0.1} & $\psi_k$ [\%] & 66 & 6.7 & 27.3 & 66 & 34\\
        &$\pi_k$ [\euro{}] & 218.14 &  22.17 &  90.2 & 218.17 & 112.37\\
        \hline
        \multirow{ 2}{*}{0.75} & $\psi_k$ [\%] & 63.3 & 7.5  & 29.2 &  63.3 & 36.7\\
        &$\pi_k$ [\euro{}] & 354.41 & 42.01 & 163.72 & 354.41 & 205.73\\
        \hline
    \end{tabular}
\end{table}

\subsection{Online regression market case}

\subsubsection{Case 1: Recursive Least Squares with an ARX model}

\begin{figure}[!ht]
    \centering
    \begin{subfigure}[b]{.48\textwidth}
    \includegraphics[width=\linewidth]{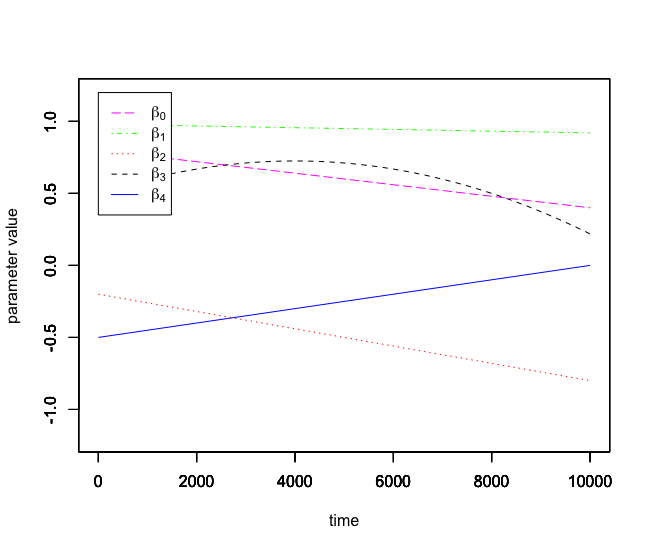}
    \caption{True parameters. \label{fig:rlsparams-true}}
\end{subfigure}    
\hfill
\begin{subfigure}[b]{.48\textwidth}
    \includegraphics[width=\linewidth]{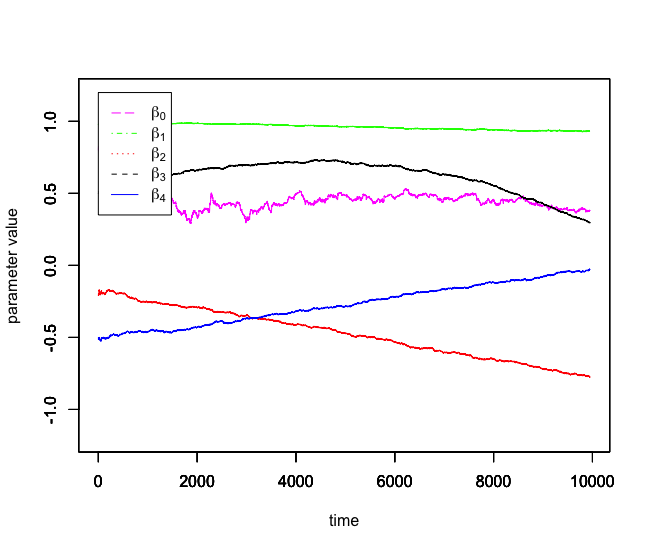}
    \caption{Estimated parameters. \label{fig:rlsparams-est}}
    \end{subfigure} 
    \caption{Temporal evolution of ARX model parameters over the period considered. \label{fig:rlsparams}}
\end{figure}

We use as a basis the same underlying model as in~\eqref{eq:qrexs} and with the same agent setup. The central agent aims at using online learning with a quadratic loss for that ARX model and with a willingness to pay of $\phi_1$=0.1\euro{} per time instant and per unit improvement in the quadratic loss function. The major difference here is that the parameters vary in time, i.e.,
\begin{equation} \label{eq:rlsexs}
    Y_t = \beta_0 + \beta_{1,t} y_{t-1} + \beta_{2,t} x_{2,t-1} + \beta_{3,t} x_{3,t-1} + \beta_{4,t} x_{4,t-1} +  \varepsilon_t \, , 
\end{equation}
as illustrated in Figure~\ref{fig:rlsparams-true}. 

\begin{figure}[!ht]
    \centering
    \begin{subfigure}[b]{.48\textwidth}
    \includegraphics[width=\linewidth]{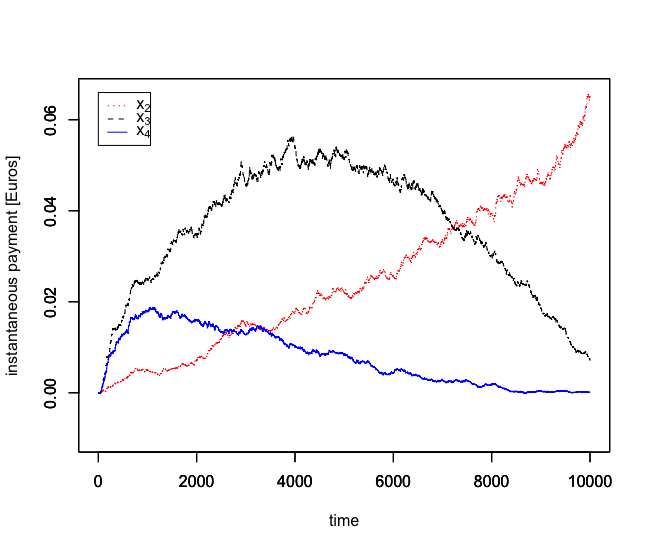}
    \caption{Instantaneous payments. \label{fig:rlsinstant}}
\end{subfigure}    
\hfill
\begin{subfigure}[b]{.48\textwidth}
    \includegraphics[width=\linewidth]{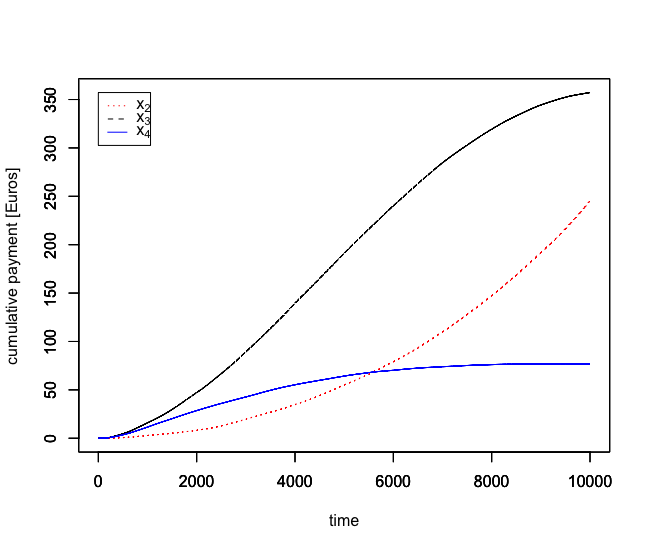}
    \caption{Cumulative payments. \label{fig:rlscum}}
    \end{subfigure} 
    \caption{Temporal evolution of the payments (instantaneous and cumulative) for the various features of the support agents over the period considered. \label{fig:rlspayments}}
\end{figure}

The central agent posts the task on the analytics platform, with online learning over a period of $T=10 000$ times steps, and defines a forgetting factor of 0.998. Since the online regression market relies on an online learning component, the parameters are tracked in time (see Figure~\ref{fig:rlsparams-est}), and with the payments varying accordingly. Such online learning schemes are very efficient in tracking parameters in the types of regression models considered here, i.e., linear in their parameters and with parameters changing smoothly in time. The payments made for the 3 features of the support agents ($x_2$ for $a_2$, as well as $x_3$ and $x_4$ for $a_3$) are depicted in Figure~\ref{fig:rlspayments}, both in terms of instantaneous payments, and cumulative ones over the period. \textcolor{black}{Since the model parameters (and their estimates) vary in time, the contributions of the various features to the improvement in the loss function also vary accordingly. This is reflected by the temporal evolution of the instantaneous payments. Here, for instance, since the estimated parameter $\hat{\beta}_4$ is getting closer to 0 as time passes, its relative importance is decreasing. In contrast, the estimate $\hat{\beta}_3$ is going up and down, and this yields a similar trajectory of the related instantaneous payments for the feature $x_3$. Finally as the importance of $x_2$ grows with time (since, even if $\hat{\beta}_2$ is negative, the contribution of $x_2$ to explaining the variance in the response variable increases), one observes a sharp rise in the corresponding instantaneous payment. Evidently, since cumulative in nature, the cumulative payments are non-decreasing with time (they can only increase or reach a plateau).}

\subsubsection{Case 2: Online learning in a quantile regression model}

\begin{figure}[!ht]
    \centering
    \includegraphics[width=0.6\textwidth]{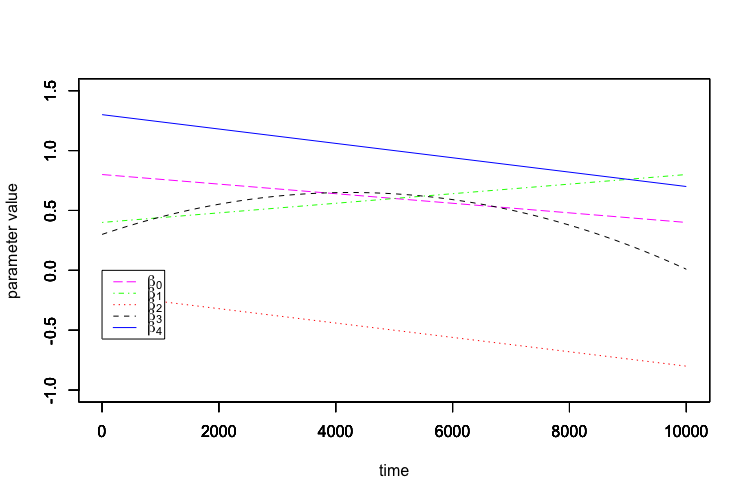}
    \caption{Temporal evolution of the process parameters over the period considered. \label{fig:rqrparams}}
\end{figure}

For this last simulation case, let us consider a linear quantile regression model, hence with a central agent aiming to perform online learning with a smooth quantile loss function. The underlying model for the process is such that
\begin{equation}
    Y_t = \beta_0 + \beta_1 x_{1,t} + \beta_2 x_{2,t} + \beta_3 x_{3,t} + \beta_4 x_{4,t} \varepsilon_t \, ,
\end{equation}
where $x_{1,t}$, $x_{2,t}$ and $x_{3,t}$ are sampled from a standard Gaussian $\mathcal{N}(0,1)$, $x_{4,t}$ is sampled from $\mathcal{U}[0.5,1.5]$ and the noise term $\varepsilon_t$ is sampled from $\mathcal{N}(0,0.3)$. It should be noted that the standard deviation of the noise is then scaled by $\beta_4 x_{4,t}$. Thinking about the distribution of $Y_t$, that means that $x_{1,t}$, $x_{2,t}$ and $x_{3,t}$ are important features to model its mean (or median), while $x_{4,t}$ will have an increased importance when aiming to model quantiles that are further away from the median (i.e., with nominal levels going towards 0 and 1). The temporal variation of the true model parameters are depicted in Figure~\ref{fig:rqrparams}.

\begin{figure}[!ht]
    \centering
    \begin{subfigure}[b]{.48\textwidth}
    \includegraphics[width=\linewidth]{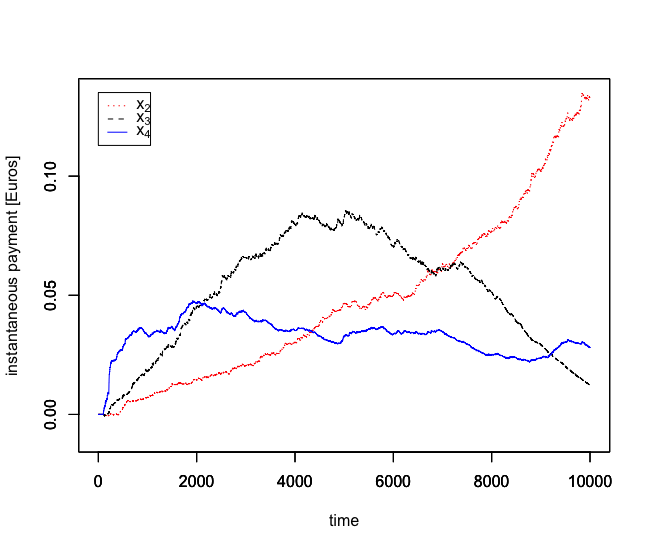}
    \caption{Instantaneous payments. \label{fig:rqrinstant}}
\end{subfigure}    
\hfill
\begin{subfigure}[b]{.48\textwidth}
    \includegraphics[width=\linewidth]{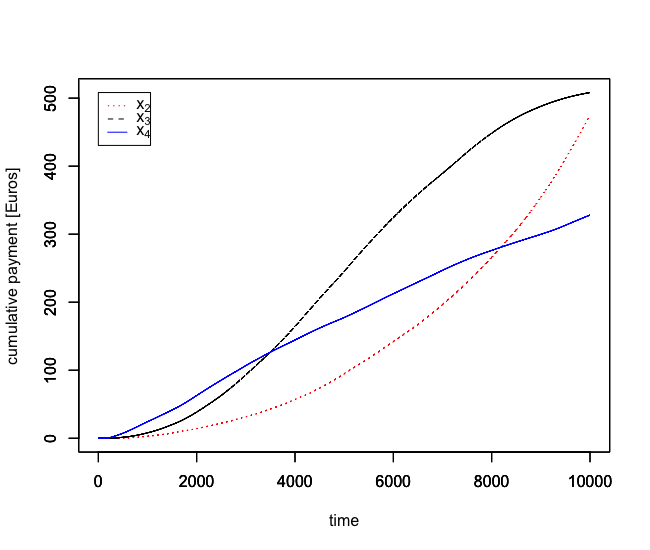}
    \caption{Cumulative payments. \label{fig:rqrcum}}
    \end{subfigure} 
    \caption{Temporal evolution of the payments (instantaneous and cumulative) for the various features of the support agents over the period considered. These results are for a nominal level $\tau=0.9$. \label{fig:rqrpayments}}
\end{figure}

The central agent posts the task on the analytics platform, with online learning over a period of $T=10 000$ time steps, and defines a forgetting factor of 0.999. The parameter $\alpha$ of the smooth quantile loss function is set to $\alpha=0.2$. The payments made for the 3 features of the support agents ($x_2$ for $a_2$, as well as $x_3$ and $x_4$ for $a_3$) are depicted in Figure~\ref{fig:rqrpayments}, both in terms of instantaneous payments, and cumulative ones over the period. These are for a choice of a nominal level of $\tau=0.9$ for the quantile of interest.

To illustrate the previous points made such that the relative value of the various features may depend on the nominal level $\tau$, Table~\ref{tab:rqrpayments} gathers the payments obtained per feature and per agent in the cases of focusing on quantiles with nominal levels 0.1, 0.25, 0.5, 0.75 and 0.9. Particularly one retrieves the fact that the payment for feature $x_4$ is 0 when looking at the median. This is in line with the definition of the data generation process, for which $x_4$ is only supposed to have value to model and predict quantiles away from the median.

\begin{table}[!ht]
    \caption{Final payments $\pi_k$ in \euro{} after $T=10 000$ time steps in the online regression market, as a function of the nominal level of the quantile of interest.}
    \label{tab:rqrpayments}
    \vspace{2mm}
    \centering
    \begin{tabular}{|c|c c c|c c|}
    \hline
        $\tau$ &$x_2$ & $x_3$ & $x_4$ & $a_2$ & $a_3$\\
        \hline
        0.1  & 712.82 & 705.71 & 332.78 & 712.82 & 1038.49 \\
        0.25 & 751.19 & 748.03 & 112.81 & 751.19 & 860.84\\
        0.5  & 747.63 & 749.59 & 0      & 747.63 & 749.59\\
        0.75 & 658.78 & 666.06 & 150.21 & 658.78 & 816.27 \\
        0.9  & 519.37 & 531.04 & 341.72 & 519.37 & 872.76\\
        \hline
    \end{tabular}
\end{table}


\section{Application to real-world forecasting problems}
\label{sec:appl}

The regression market approach we proposed is originally developed with energy forecasting applications in mind. Besides the simulation-based case studies considered in the above to illustrate the workings and applicability of regression markets, we focus here on real-world applications using data from South Carolina (USA). Regression models are used as a basis for forecasting, hence with a learning stage (batch and online) and an out-of-sample stage (for genuine forecasting). We restrict ourselves to a fairly simple setup with 1-hour ahead forecasting, though other lead times could be similarly considered (possibly requiring different input data and regression models). The aim is certainly not to develop a forecasting approach which is to be better than the state-of-art, but to show how our regression market mechanism {\it (i)} incentivizes data sharing, {\it (ii)} yields improved forecasts, and {\it (iii)} appropriately compensates support agents for their contribution to improvement in the loss function (and the forecasts) of the central agent.

\subsection{Data description and modeling setup}

To ensure that the application to real-world data can be reproduced and comprises a good starting point for others, we use a dataset from an open database for renewable energy data in the USA. The wind power generation data for a set of 9 wind farms in South Carolina (USA) was extracted from the Wind Integration National Dataset (WIND) Toolkit described in \citet{Draxl2015}. The data is hence not completely real, but still very realistic in capturing the local and spatio-temporal dynamics of wind power generation within an area of interest. It is owing to such spatio-temporal dynamics that one expects to see benefits in using others' data to improve power forecasts -- see \citet{Cavalcante2017} and \citet{Messner2019} for instance. An overview of the wind farms and of their characteristics is given in Table~\ref{tab:sites}. These are all within 150 kms of each other. Wind power measurements are available for a period of 7 years, from 2007 to 2013, with an hourly resolution. For the purpose of the regression and forecasting tasks, all power measurements are normalized and hence take values in $[0,1]$. An advantage of this type of data is that there is no missing and no suspicious data point to be analyzed and possibly to be removed. In this setup, each wind farm may be seen as an agent. We therefore have 9 agents $a_1, \hdots, a_9$ who can take the role of either central or support agent. Let us write $y_{j,t}$ the power measurement of agent $a_j$ at time $t$, which is a realization of the random variable $Y_{j,t}$.

\begin{table}[!ht]
\caption{Sites considered in South Carolina, USA, with data available for a period of 7 years (2007-2013). Notations: C$_\text{f}$ for capacity factor, P$_\text{n}$ for nominal capacity. The ``id" is that from the Wind Toolkit database. \label{tab:sites}}
\centering
\begin{tabular}{|c|c|ccccc|}
\hline
Agent & id & C$_\text{f}$ [\%] & P$_\text{n}$ [MW]	& Lat./Long. & County & Elevation [m]  \\
\hline
$a_1$ & 4456 &	34.11 &	1.75  &	34.248/-79.75  & Florence & 36.17 \\
$a_2$ & 4754 &	35.75 &	2.96  &	34.02/-79.537 & Florence & 17.5 \\
$a_3$ & 4934 &  36.21 &	3.38  &	33.925/-79.958 & Florence & 36.2 \\
$a_4$ & 4090 &  26.6  &	16.11 &	34.732/-82.122 & Laurens & 219.73 \\
$a_5$ & 4341 &  28.47 &	37.98 &	34.556/-81.889 & Laurens & 182.31 \\
$a_6$ & 4715 &  27.37 &	30.06 &	34.334/-82.133 & Laurens & 164.07 \\
$a_7$ & 5730 &	34.23 &	2.53  &	33.136/-80.857 & Colleton & 42.75 \\
$a_7$ & 5733 &	34.41 &	2.6	  & 33.112/-80.665 & Colleton & 27.0 \\
$a_9$ & 5947 &	34.67 &	1.24  &	32.641/-80.504 & Colleton & 5.4 \\
\hline
\end{tabular}
\end{table}

Emphasis is placed on very short-term forecasting (i.e., 1 hour ahead) as a basis for illustration of regression markets for a real-world setup. This allows us to use fairly simple time-series modeling and forecasting approaches. Those may readily extended to the case of further lead time, possibly using additional input features, e.g., from remote sensing and weather forecasts. More advanced modeling approaches could additionally be employed, e.g., if aiming to account for the nonlinearity and double-bounded nature of wind power generation \citep{Pinson2012}.

For a given central agent $a_i$ and support agents $a_j, \, j \neq i$, the basic underlying model considered for the regression markets writes
\begin{equation}
    Y_{i,t} \, = \, \beta_0 \, + \, \sum_{\delta=1}^\Delta y_{i,t-\delta} \, + \, \sum_{j \neq i} \sum_{\delta=1}^\Delta y_{j,t-\delta} \, + \, \varepsilon_{i,t} \, ,
\end{equation}
which is simply an ARX model with maximum lag $\Delta$. In principle, one would run a data analysis exercise to pick the number of lags, or alternatively cross-validation. We assume here that expert knowledge, or such an analysis, allowed to conclude for the use of 2 lags for the central agent, and 1 lag only for the support agents.

For both cases in the following, we place ourselves within a simplified electricity market setup, where it is assumed that wind farms have to commit to a scheduled power generation 1-hour ahead. They then get a set price per MWh scheduled (e.g., 40\$), though with a penalization afterwards for deviation from schedule. This penalization is proportional to a chosen loss function. In the first case, for the batch and out-of-sample regression markets, a quadratic loss function is used. This translates to the agents assessing their forecasts in terms of Mean Square Error (MSE) and aiming to reduce it. In the second case, we envision an asymmetric loss as in European electricity markets (with 2-price imbalance settlement), where agents then aim to reduce a quantile loss, with the nominal level $\tau$ of the quantile being a direct function of the asymmetry between penalties for over- and under-production \citep{Morales2014}. In both cases, agents could perform an analysis to assess the value of forecasts in those markets, as well as their willingness to pay to improve either quadratic of quantile loss. Here, we consider that all agents have valued their willingness to pay, denoted $\phi$ and expressed in \$ per percent point improvement in their loss function and per data point, to be shared between in-sample (batch or online) and out-of-sample regression markets. We use percent point improvement as those loss functions are normalized.

\subsection{Batch and out-of-sample regression markets}

In the batch and out-of-sample case, the first 10 000 time instants (so, a bit more than a year) are used to train the regression models within the batch regression market, while the following 10 000 time instants are for the out-of sample forecasting period, hence for the out-of-sample regression market. 

Let us first zoom in on the case of agent $a_1$, splitting her willingness to pay as $\phi=$ 0.5\$ per percent point improvement in quadratic loss and per data point within the batch regression market, and $\phi=$ 1.5\$ for the out-of-sample regression market. In that case, in-sample through the batch regression market, the quadratic loss is reduced from 2.82\% of nominal capacity to 2.32\% thanks to the data of the support agents. And, out-of-sample, that loss decreases from 3.09\% to 2.53\% when relying on the support agents. The allocation policies $\psi_j$ as well as payments $\pi_j$ are gathered in Table~\ref{tab:batcha1}. The overall payment of central agent $a_1$ for the two markets is of 10 855.98\$. \textcolor{black}{As mentioned when introducing regression markets, there may obviously be disparities between the value of features and data of support agents at the batch and out-of-sample stages. It is clear here for instance if looking at the Shapley allocation for support agent $a_3$, where the in-sample allocation is of 32.35\% and then dropping to 22.27\% out-of-sample. For all other support agents, the Shapley allocation values increase when going from batch to out-of-sample regression markets, somewhat compensating for the substantial change observed for $a_3$.}

\begin{table}[!ht]
\caption{Payments $\pi_j$ and Shapley allocation policies $\psi_j$ in both batch and out-of-sample regression markets, with $a_1$ being the central agent and all others being support agents. \label{tab:batcha1}}
\centering
\begin{tabular}{|c|c|cccccccc|}
\hline
Market & & $a_2$& $a_3$& $a_4$& $a_5$& $a_6$& $a_7$& $a_8$& $a_9$\\
\hline
\multirow{ 2}{*}{batch} & $\psi_j$ [\%] &	23.17 & 32.35 & 9.75 & 8.24 & 7.92 & 6.36 & 6.84 & 3.81 \\
                        & $\pi_j$  [\$]  & 574.29 & 801.72 & 241.52 & 204.12 & 196.19 & 157.71 & 169.42 & 94.38 \\
\multirow{ 2}{*}{out-of-sample} & $\psi_j$ [\%] & 26.96 & 22.27 & 10.74 & 11.47 & 9.33 & 7.15 & 7.07 & 4.34 \\
                        & $\pi_j$ [\$]   & 2284.52 & 1887.31 & 909.83 & 972.02 & 790.49 & 605.65 & 599.15 & 367.66 \\
                        \hline
\multicolumn{2}{|c|}{Total payment [\$]} & 2858.81 & 2689.03 & 1151.35 & 1176.14 &  986.68 &  763.36 &  768.57 &  462.04 \\
\hline
\end{tabular}
\end{table}

First of all, agents $a_2$ and $a_3$ provide the features that make the strongest contribution towards lowering the quadratic loss, both in-sample in the batch regression market and for genuine forecasting through the out-of-sample regression market. However, one of them ($a_3$) has higher Shapley allocation policy values in-sample, and the other one ($a_2$) out-of-sample. It is then reflected by the payments. Eventually, from the perspective of the support agents, those total payments should be divided by 20 000, to reflect the unit value of each data point provided for their features. For instance, the value of an individual data point of $a_2$ is of 14\textcent, and of only 2.3\textcent{} for $a_9$.

\begin{figure}[!ht]
    \centering
    \begin{subfigure}[b]{.48\textwidth}
    \includegraphics[width=\linewidth]{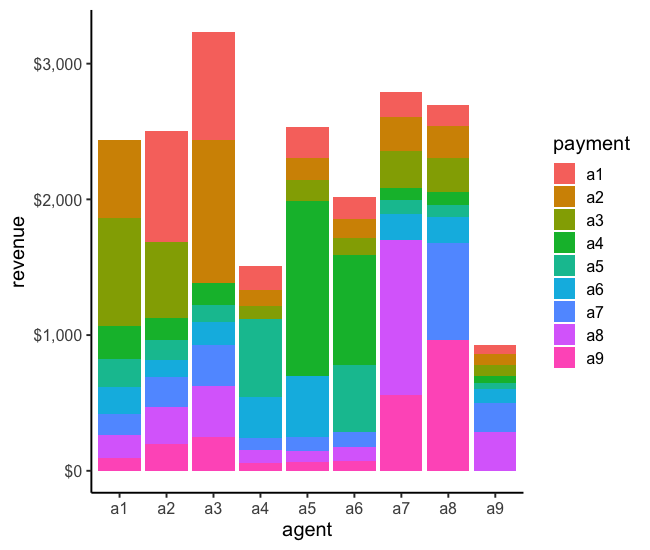}
    \caption{Batch regression market. \label{fig:SCbatchrevenues}}
\end{subfigure}    
\hfill
\begin{subfigure}[b]{.48\textwidth}
    \includegraphics[width=\linewidth]{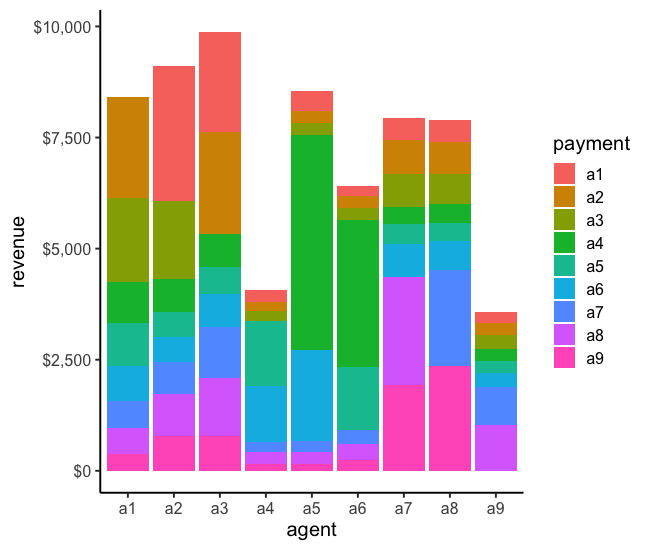}
    \caption{Out-of-sample regression market. \label{fig:SCbatchoutrevenues}}
    \end{subfigure} 
    \caption{Cumulative revenues of all agents in both batch and out-of-sample regression markets. \label{fig:SCbatchandoutrevenues}}
\end{figure}

Since we have 9 agents in this South Carolina case study, they can all play the role of the central agent, and use data from other agents to improve their forecasts. This means, for instance, that eventually the revenue of $a_9$ comes from parallel regression markets where agents $a_1, \hdots, a_8$ play the role of central agent and pay $a_9$ for her data. For simplicity, we rely on the same setup and willingness to pay for all agents. The cumulative revenues of the 9 agents are depicted in Figure \ref{fig:SCbatchandoutrevenues}, for both batch and out-of-sample regression markets. The value of the data of the different agents varies significantly depending on the central agent considered. As an example, the data of $a_1$ is highly valuable to agents $a_2$ and $a_3$ both in batch and out-of-sample regression markets, but not so much to the other agents. The heterogeneity of those payments and revenues certainly reflects the geographical positioning and prevailing weather conditions in this area of South Carolina. Looking at the cumulative revenues for all agents, it is also clear that the data of agents $a_4$ and $a_9$ carries much less value overall than the data of the others. For instance for the out-of-sample regression market (over a period of 10 000 time instants), by providing data to all other agents, the unit value of a single data point of the agents vary from 46\textcent{} for $a_9$ to 99\textcent{} for $a_3$. \textcolor{black}{In  the batch regression case, the in-sample and out-of-sample assessment of the loss function and resulting Shapley allocation policies may be fairly different, since based on different time periods and since quality of model fitting may not always be reflective of genuine contribution to forecast quality. This is observed here based on the differences in payment and revenues for the batch and out-of-sample regression markets.}

The payments of a central agent towards support agents is proportional to forecast improvements in terms of a quadratic loss. The normalized MSE of 1-step ahead forecasts (score consistent with the quadratic loss) are gathered in Table~\ref{tab:MSEbatchoos}, over both batch learning and out-of-sample forecasting periods. As expected, the normalized MSE values are always lower when the agents have used the regression markets since, if there were no improvement in terms of a quadratic loss, there would be no payment to support agents.

\begin{table}[!ht]
\caption{Normalized MSE for all agents (expressed in \% of nominal capacity), during both batch learning and out-of-sample forecasting phases, also with and without the use of data from the support agents. \label{tab:MSEbatchoos}}
\centering
\begin{tabular}{|c|c|ccccccccc|}
\hline
 & & $a_1$& $a_2$& $a_3$& $a_4$& $a_5$& $a_6$& $a_7$& $a_8$& $a_9$\\
\hline
\multirow{ 2}{*}{batch} & without &	2.82 & 2.90 & 2.88 & 3.51 & 3.35 & 3.19 & 2.76 & 2.86 & 2.33\\
                        & with    & 2.32 & 2.38 & 2.22 & 3.18 & 2.83 & 2.78 & 2.20 & 2.31 & 2.12\\
                        \hline
\multirow{ 2}{*}{out-of-sample} & without & 3.09 & 2.78 & 3.12 & 3.48 & 3.33 & 3.27 & 3.05 & 3.13 & 2.78 \\
                        & with   & 2.53 & 2.51 & 2.44 & 3.19 & 2.75 & 2.84 & 2.51 & 2.60 & 2.52\\
                        \hline 
\end{tabular}
\end{table}

\subsection{Online and out-of-sample regression markets}

In the online case, we do not have a clear separation between the batch learning and out-of-sample forecasting periods, in the sense that at each time instant $t$, when new data becomes available, one may assess the forecast issued at time $t-1$ for time $t$ (for the out-of-sample regression market), and in parallel update the parameter estimates for the regression model through the online regression market. Then, a new forecast (for time $t+1$) is issued.

We consider here a setup that is similar to the batch case in the above, i.e., with a willingness to pay of the agent split between the online regression market ($\phi=0.2$\$ per percent point improvement in the loss function and per data point) and the out-of-sample regression market ($\phi=0.8$\$ per percent point improvement in the loss function and per data point). Instead of the quadratic loss function, emphasis is placed on quantile regression instead, hence using the smooth quantile loss. We arbitrarily choose the nominal level of the quantile to be $\tau=0.55$, to reflect the asymmetry of penalties in an electricity market with 2-price imbalance settlement at the balancing stage. This corresponds to the case of an electricity market that penalizes wind power producers slightly more for over-production than for under-production. The smoothing parameter for the smooth quantile loss is set to $\alpha=0.2$, while the forgetting factor is set to $\lambda=0.995$. Note that these are not optimized parameters. These could be optimized through cross-validation for instance.

In contrast to the batch and quadratic loss case, not all agents' features may be valuable. We use a screening approach: if the Shapley allocation policies values are negative after the burn-in period, those agents are removed. The burn-in period is based on the first 500 time instants.

\begin{figure}[!ht]
    \centering
    \begin{subfigure}[b]{.48\textwidth}
    \includegraphics[width=\linewidth]{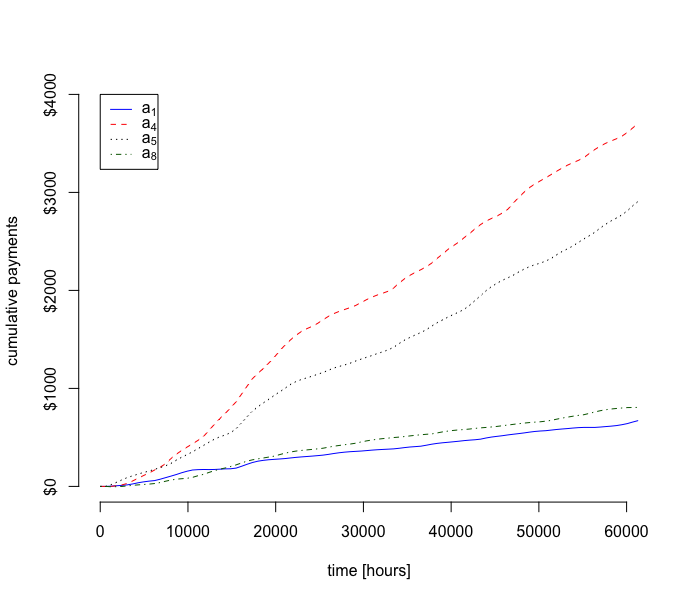}
    \caption{Online regression market. \label{fig:SConlinecumpaymentsa6}}
\end{subfigure}    
\hfill
\begin{subfigure}[b]{.48\textwidth}
    \includegraphics[width=\linewidth]{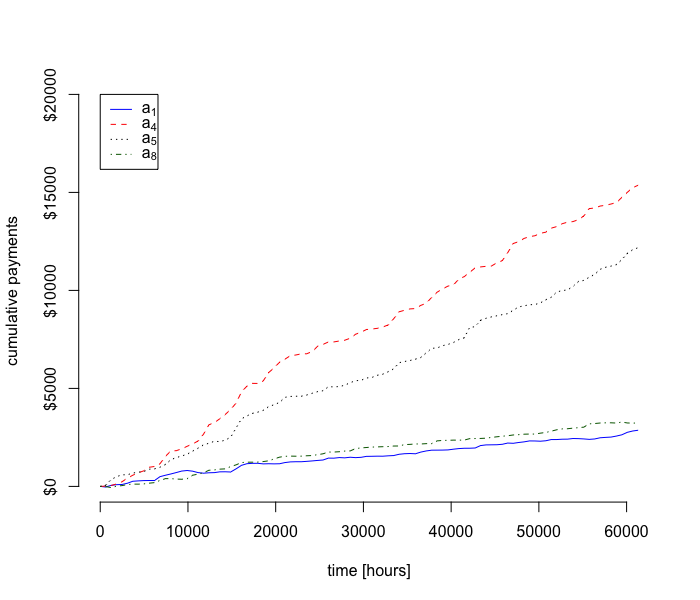}
    \caption{Out-of-sample regression market. \label{fig:SCooscumpaymentsa6}}
    \end{subfigure} 
    \caption{Evolution of the cumulative payments of $a_6$ towards agents $a_1$, $a_4$, $a_5$ and $a_8$, in both online and out-of-sample regression markets, over a period of 7 years. \label{fig:SCcumpaymentsa6}}
\end{figure}

Let us first concentrate on agent $a_6$ for instance, who, after the burn-in period, only uses data from agents $a_1$, $a_4$, $a_5$ and $a_8$. The cumulative payments of $a_6$ to these agents are depicted in Figure~\ref{fig:SCcumpaymentsa6} as a function of time, for both online and out-of-sample regression markets. Clearly, $a_4$ and $a_5$ receive significantly higher payments than the other two agents. Also, there are periods with higher and lower payments, since these cumulative payment lines are not straight lines. Over the first 1.5 years (app. 13 000 hours) the data from $a_1$ leads to higher payments than the data from $a_8$, while it is the opposite situation for the remaining 5.5 years.

\begin{figure}[!ht]
    \centering
    \begin{subfigure}[b]{.48\textwidth}
    \includegraphics[width=\linewidth]{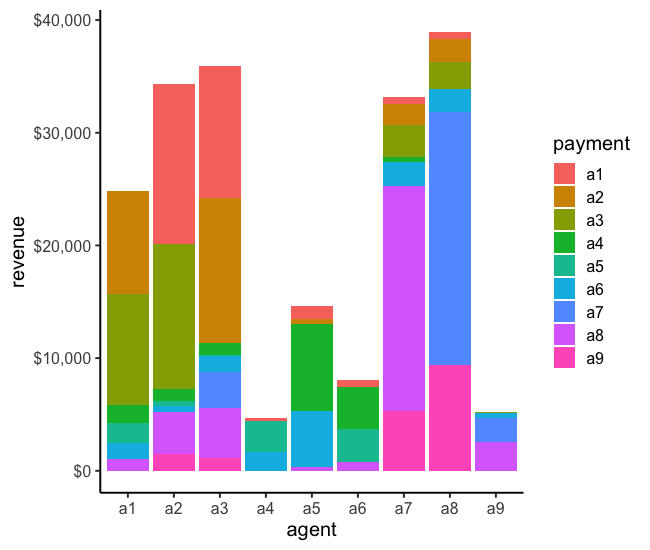}
    \caption{Online regression market. \label{fig:SConlinerevenues}}
\end{subfigure}    
\hfill
\begin{subfigure}[b]{.48\textwidth}
    \includegraphics[width=\linewidth]{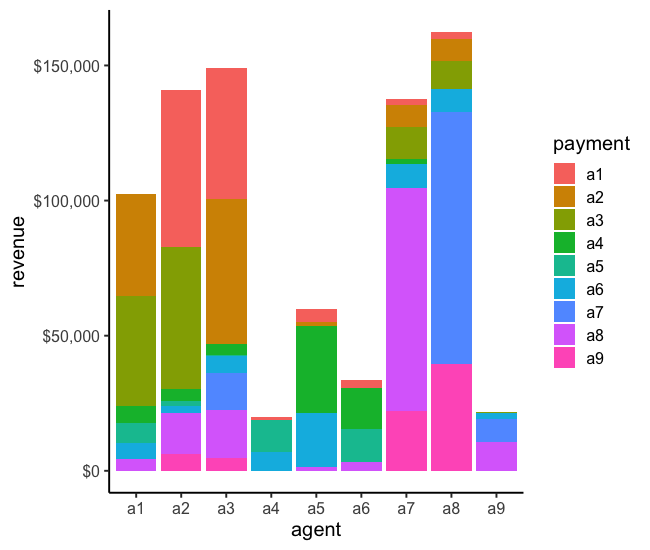}
    \caption{Out-of-sample regression market. \label{fig:SConlineoutrevenues}}
    \end{subfigure} 
    \caption{Final cumulative revenues of all agents in both online and out-of-sample regression markets, after 7 years. \label{fig:SConlineandoutrevenues}}
\end{figure}

Finally, we perform the same study for all agents acting as central agents, and aiming to improve their quantile forecasts based on the data of others. They engage in both online and out-of-sample regression markets, under the exact sames conditions (i.e., model, willingness to pay, hyperparameters, etc.). The overall revenues obtained after the 7-year period are depicted in Figure~\ref{fig:SConlineandoutrevenues}, for both regression markets. The differences in the value of the data of the various agents is even higher than in the batch case with quadratic loss function. Certain agents like $a_4$, $a_6$ and $a_9$ receive payments from 3 or 4 other agents only, and with much lower revenues overall. And, while $a_3$ was the agent who obtained the highest revenue in the previous study, it is now $a_8$ who obtains the highest revenue.

There are also some consistent results with the previous case, for instance with $a_1$ giving large payments to $a_2$ and $a_3$, as well as $a_7$ receiving large payments from $a_8$. For the agents that have the most valuable data, the overall revenues over the 7-year period are quite sizeable, for instance reaching 200 000\$ for $a_8$. This represents a unit value of 3.26\$ per data point being shared with the other agents.

\textcolor{black}{Interestingly, one can observe from Figure~\ref{fig:SConlineandoutrevenues} that the distribution of revenues and payments is very similar between the in-sample and out-of-sample regression market cases. This is in contrast with what was observed for the batch. This can be explained by the fact that, in an online learning framework, the same forecast errors are iteratively used for {\it (i)} out-of-sample assessment of loss function and Shapley allocation policies, and  {\it (ii)}  in-sample assessment within the recursive updates of model parameters. These two assessments and related Shapley allocations are very close to each other since the time-varying loss estimates in online learning, for instance at time $t$, are very close estimates of the forecast accuracy to be expected when issuing a forecast at that time.}

\section{Conclusions and perspectives}
\label{sec:concl}

The digitalization of energy system has brought in a lot of opportunities towards improving the operations of energy systems with increased penetration of renewable energy generation, decentralization and more proactive demand, liberalization of energy markets, etc. For many operational problems, it is assumed that data can be shared and centralized for the purpose of solving the analytics task at hand. However, in practice, it is rarely the case that the agents are willing to freely share their data. With that context in mind, we have proposed here a regression market approach, which may allow to incentivize and reward data sharing for one family of analytics task, regression, for instance widely used as a basis for energy forecasting. Obviously, in the future, the concepts and key elements of the approach should be extended to the case of other analytics tasks, e.g. classification, filtering, etc., and to the nonlinear case. In addition, the properties of the various regression markets may be further studied, for instance in a regret analysis framework, to provide some interesting bounds and potential fairness implications.

Mechanism design for data and information has specifics that differ from the case of considering other types of commodities. For instance, the value of information carried by data is a function of the analytics task at hand, timeliness in the data sharing, possibly data quality, among other aspects. Therefore, this triggers the need to rethink some of the basic concepts of mechanism design within that context. Importantly, even with a mechanism exhibiting desirable properties being into place, it may be difficult for all agents involved to assess their willingness to buy and willingness to sell. On the buying side, this quantification most likely relies on a decision process and a related loss function. However, if different decision processes are intertwined and possibly in a sequential manner, that willingness to pay might be more difficult to reveal. On the selling side, the willingness to sell may be affected by actual cost of obtaining the data (as well as storing and sharing it), plus possibly privacy-related and competition-related aspects. Indeed, imagining the case of renewable energy producers all participating in the same electricity market, sharing data could eventually affect an existing competitive advantage, by making other market participants more competitive. From an overall societal perspective, one would expect increased social welfare though, since such mechanism would allow for making optimal use of all available information.

\section*{Acknowledgments}

The research leading to this work is being carried out as a part of the Smart4RES project (European Union’s Horizon 2020, No. 864337). The sole responsibility of this publication lies with the authors. The European Union is not responsible for any use that may be made of the information contained therein. The authors are indebted to the developers and authors of the Wind Toolkit, for making such data available. The authors are also grateful for the comments and suggestions provided by the reviewers and editor who handled the paper, which allowed to improve the paper. Finally, Ricardo Bessa and Carla Goncalves at INESC Porto are to be acknowledged for numerous and fruitful discussions related to data markets for energy system applications.

\section*{Author contributions}

All authors contributed to the study conception and design. The development of the methodology part, simulation and case-study applications, were performed by Pierre Pinson. The first draft of the manuscript was written by Pierre Pinson and all authors commented on previous versions of the manuscript. All authors read and approved the final manuscript, while contributing to the revisions.

\section*{Statements and declarations}

The authors declare no conflict of interest.

\appendix 

\section{Proof of Theorem 1}
\label{sec:proofoftheorem1}

Let us give a proof in the following for all the properties covered in Theorem 1, on a point by point basis.

\vspace{2mm}
\noindent
(i) {\bf Budget balance}

\vspace{1mm}
A property of the Shapley allocation policies is that they are balanced, i.e.,  whatever the regression model, loss function $l$ and batch of data used for estimation, one has 
\begin{equation}
    \sum_k \psi_k(l) = 1 \, . \nonumber
\end{equation}
Consequently,
\begin{eqnarray}
   \sum_k \pi_k &=& \sum_k T (L_{\omega_i}^*- L_\Omega^*) \phi_i \psi_k (l) \nonumber \\
   &=& T (L_{\omega_i}^*- L_\Omega^*) \phi_i \sum_k \psi_k(l) \nonumber \\
   &=&  T (L_{\omega_i}^*- L_\Omega^*) \phi_i \, . \nonumber
\end{eqnarray}
Hence, the sum of the revenues of the support agents is equal to the payment of the central agent.

\vspace{2mm}
\noindent
(ii) {\bf Symmetry}

\vspace{1mm}
Assume that 2 support agents have identical features $x_k$ and $x_k'$. This would then imply that
\begin{equation}
    S^*_{\omega_i \cup \omega \cup k} = S^*_{\omega_i \cup \omega \cup k'} \, , \quad \forall \omega \in \Omega_j \setminus k,k' \, . \nonumber
\end{equation}
One can therefore deduce that these two features will have the same Shapley allocation policy, i.e., $\psi_k (l) = \psi_{k'} (l)$. In view of the payment definition in \eqref{eq:paymentbatch}, they will also receive the same payment, $\pi_k = \pi_{k'}$. It also means that any permutation of indices will yield the same payments.

\vspace{2mm}
\noindent
(iii) {\bf Truthfulness}, i.e., support agents only receive their maximum potential revenues when reporting their true feature data

\vspace{1mm}
We consider here models that are linear in their parameters. Fundamentally, the estimation problem boils down to
\begin{equation} \label{eq:regestbase}
   \hat{\boldsymbol{\beta}}_{\omega} = \argmin_{\boldsymbol{\beta}_{\omega}} \, \mathbb{E} \left[ l \left( Y_t-\boldsymbol{\beta}_{\omega}^\top \tilde{\mathbf{x}}_t \right)\right] \, ,
\end{equation}
where the expectation is eventually replaced by the batch in-sample estimator in~\eqref{eq:batchcentral2}. In the case one of the support agents does not truthfully report data, the data that enters the estimation problem is $\tilde{\mathbf{x}}_t+\eta_t, \, \forall t$ (where the noise only affects the feature of that support agent). If $\eta_t$ is a constant, the solution of~\eqref{eq:batchcentral} is not affected, hence the support agent cannot obtain increased revenues. If instead $\eta_t$ is a centred noise with finite variance, one would solve instead
\begin{equation} \label{eq:batchnoise}
    \argmin_{\boldsymbol{\beta}_{\omega}} \, \mathbb{E} \left[ l \left( Y_t-\boldsymbol{\beta}_{\omega}^\top ( \tilde{\mathbf{x}}_t + \eta_t) \right) \right] \, ,
\end{equation}
which will yield a vector of model parameters $\hat{\boldsymbol{\beta}}_{\omega} + \delta \hat{\boldsymbol{\beta}}_{\omega}$ that is different from $\hat{\boldsymbol{\beta}}_{\omega}$. The expected loss function at that point can be written as
\begin{align}
    \mathbb{E} \left[ l \left( \hat{\boldsymbol{\beta}}_{\omega} + \delta \hat{\boldsymbol{\beta}}_{\omega} \right) \right] & =  \mathbb{E} \left[ l \left( Y_t- (\hat{\boldsymbol{\beta}}_{\omega} + \delta \hat{\boldsymbol{\beta}}_{\omega})^\top ( \tilde{\mathbf{x}}_t + \eta_t)  \right) \right] \nonumber \\
    & = \mathbb{E} \left[ l \left( Y_t - (\hat{\boldsymbol{\beta}}_{\omega} + \delta \hat{\boldsymbol{\beta}}_{\omega})^\top \tilde{\mathbf{x}}_t - (\hat{\boldsymbol{\beta}}_{\omega} + \delta \hat{\boldsymbol{\beta}}_{\omega})^\top  \eta_t \right) \right] \, . \nonumber
\end{align}
Since the expectation of a convex function is a convex function and $(\hat{\boldsymbol{\beta}}_{\omega} + \delta \hat{\boldsymbol{\beta}}_{\omega})^\top  \eta_t$ is a noise term, one has 
\begin{equation}
    \mathbb{E} \left[ l \left( Y_t - (\hat{\boldsymbol{\beta}}_{\omega} + \delta \hat{\boldsymbol{\beta}}_{\omega})^\top \tilde{\mathbf{x}}_t - (\hat{\boldsymbol{\beta}}_{\omega} + \delta \hat{\boldsymbol{\beta}}_{\omega})^\top  \eta_t \right) \right] \geq \mathbb{E} \left[ l \left( Y_t - (\hat{\boldsymbol{\beta}}_{\omega} + \delta \hat{\boldsymbol{\beta}}_{\omega})^\top \tilde{\mathbf{x}}_t \right) \right]  \, . \nonumber
\end{equation}
And then, since we know that $\hat{\boldsymbol{\beta}}_{\omega}$ is the solution of~\eqref{eq:regestbase}, it follows that
\begin{equation}
    \mathbb{E} \left[ l \left( Y_t - (\hat{\boldsymbol{\beta}}_{\omega} + \delta \hat{\boldsymbol{\beta}}_{\omega})^\top \tilde{\mathbf{x}}_t \right) \right] \geq  \mathbb{E} \left[ l \left( Y_t - \hat{\boldsymbol{\beta}}_{\omega}^\top \tilde{\mathbf{x}}_t \right) \right] \, . \nonumber
\end{equation}

As a consequence, looking at the payment for feature $x_k$ based on Shapley allocation policies, 
\begin{equation} \label{eq:payshapley}
    \pi_k (l) = T \sum_{\omega \subseteq \Omega_{-i} \setminus \{x_k\}} \frac{|\omega|! (n-|\omega|-1)!}{n!} \left( L^*_{\omega_i \cup \omega} - L^*_{\omega_i \cup \omega  \cup \{x_k\}}\right) \, ,
\end{equation}
we expect that the loss function when using altered feature $x_k + \eta$ will be higher than if using the non-altered feature $x_k$. The payment will then be less (or equal). One should note, however, that this result is valid if one could use the true expected loss. In practice, only an in-sample estimator ($L_\omega$) is available and used in the payment calculation. The result may then be affected by sampling uncertainty.

\vspace{2mm}
\noindent
(iv)  {\bf Individual rationality}, i.e., the revenue of the support agents is at least 0

\vspace{1mm}
Property~\ref{prop:allocationpolicies} stipulates that $\psi_k (l) \geq 0$ (and less than 1). It readily follows from the definition of payments in~\eqref{eq:paymentbatch} and~\eqref{eq:paymentbatchagentj} that payments can only be such that $\pi_k \geq 0$ and $\pi(a_j) \geq 0$.

\vspace{2mm}
\noindent
(v) {\bf Zero-element}, i.e., a support agent that does not provide any feature, or provide a feature that has no value (in terms of improving the loss estimate $L_\omega$), gets a revenue of 0

\vspace{1mm}
In the case no feature is provided, there is obviously no payment to the support agent for that feature. In parallel, if a feature $x_k$ has no value this means that
\begin{equation}
    L_{\omega_i \cup \omega \cup\{x_k\}} - L_{\omega_i \cup \omega } = 0, \, \quad \forall \omega \subset \Omega_{-i} \, , \nonumber
\end{equation}
which hence yield $\psi_k(l)=0$, for both leave-one-out and Shapley allocation policies. Consequently, the payment is $\pi_k=0$. Note that in practice, due to sampling effect over a limited batch of data, it is highly unlikely that the value of a feature $x_k$ is exactly 0.

\vspace{2mm}
\noindent
\textcolor{black}{(iv)  {\bf Linearity}, i.e., for any two set of features $\omega$ and $\omega'$, the revenue obtained by sharing $\omega \cup \omega'$ is equal to the sum of the revenues if having shared $\omega$ and $\omega'$ separately}

\vspace{1mm}
\textcolor{black}{The linearity property of the regression markets directly comes from the linearity property of Shapley values. I.e., for any two sets of features $\omega$ and $\omega'$, in terms of Shapley allocation policies one readily has that
\begin{equation}
    \psi_{\omega \cup \omega'} = \psi_{\omega} +\psi_{\omega'}  , \nonumber
\end{equation}
which necessary implies that, in terms of payments to the support agents for the sets of features $\omega$ and $\omega'$,
\begin{equation}
    \pi_{\omega \cup \omega'} = \pi_{\omega} +\pi_{\omega'}  . \nonumber
\end{equation}
It should be noted that this property also holds with the leave-one-out allocation policies, if the input features are independent.}

\end{document}